\documentclass[usenatbib%, draft
]{mnras}
\usepackage[T1]{fontenc}
\usepackage{ae,aecompl}

\usepackage{graphicx}
\usepackage{hyperref}
\usepackage{amsmath}
\usepackage{amssymb}
\usepackage{mathtools}
\usepackage{bm}
\usepackage{cleveref}
\usepackage{tensor}
\usepackage{braket}

\newcommand{\spc}{\quad \quad \quad}
\newcommand{\paral}{\mathbin{ \, \!/\mkern-5mu/\!} \,}

\title[A relativistic correction to the mutual friction coupling time-scale]
{A universal formula for the relativistic correction to the mutual friction coupling time-scale in neutron stars}

\author[L. Gavassino, M. Antonelli, P. Pizzochero \& B. Haskell]{
    L.~Gavassino$^{1}$ \thanks{E-mail: lorenzo.gavassino@gmail.com},
    M.~Antonelli$^{1}$ \thanks{E-mail: mantonelli@camk.edu.pl}, 
    P.~M.~Pizzochero$^{2,3}$,
    B.~Haskell$^{1}$\\ \\
    $^{1}$Nicolaus Copernicus Astronomical Center of the Polish Academy of Sciences, Bartycka 18, 00-716 Warszawa, Poland \\
    $^{2}$Istituto Nazionale di Fisica Nucleare, sezione di Milano, Via Celoria 16, 20133 Milano, Italy\\
    $^{3}$Dipartimento di Fisica, Universit\`a degli Studi di Milano, Via Celoria 16, 20133 Milano, Italy}

\begin{document}
    %
    %%\date{Submitted to MNRAS in Feb-2016.}
    %
    \pagerange{\pageref{firstpage}--\pageref{lastpage}} \pubyear{2018}
    \maketitle
    \label{firstpage}

\begin{abstract}
Vortex mediated mutual friction governs the coupling between the superfluid and normal components in neutron star interiors. By, for example, comparing precise timing observations of pulsar glitches with theoretical predictions it is possible to constrain the physics in the interior of the star, but to do so an accurate model of the mutual friction coupling in General Relativity is needed. We derive such a model directly from Carter's multi-fluid formalism, and study the vortex structure and coupling timescale between the components in a relativistic star. We calculate how General Relativity modifies the shape and the density of the quantised vortices and show that, in the quasi-Schwarzschild coordinates, they can be approximated as straight lines for realistic neutron star configurations. Finally, we present a simple universal formula (given as a function of the stellar compactness alone) for the relativistic correction to the glitch rise-time, which is valid under the assumption that the superfluid reservoir is in a thin shell in the crust or in the outer core. This universal relation can be easily employed to correct, a posteriori, any Newtonian estimate for the coupling time scale, without any additional computational expense.
\end{abstract}

\begin{keywords}
Stars: neutron - Pulsars: general - Dense matter - Gravitation - Hydrodynamics
\end{keywords}

%===============================================================
% Introduction
%===============================================================

\section{Introduction}\label{sec:intro}

Pulsars are known to be among the most stable clocks in the universe, but their timing irregularities can help unveil the mystery of their interior structure. The sudden spin-up events called glitches are thought to be a manifestation of the presence of a neutron superfluid in the interior of the neutron star \citep{andersonitoh1975}. According to this theory, there is a region of the star, the exact nature and extension of which is still uncertain  \citep{andersson+2012,chamel2013}, in which the quantised vortices of the superfluid are mostly pinned to the normal component during almost the whole life of the neutron star \citep{alpar77,epsteinbaym88}. This results in the formation of an angular momentum reservoir which, when the lag between the superfluid and the normal component becomes too large, is released in a catastrophic event, producing the glitch (for a recent review see \citealt{haskellmelatos2015}).  

Many attempts to constrain the physical properties of neutron stars from observations of the relaxation process have already been made \citep{baym+1969,datta1993,LE99,graber2018ApJ}, but with the development of the Square Kilometer Array \citep{Lazio2009,SKAtuttiinsieme2018} and the Five-hundred-meter Aperture Spherical Telescope \citep{Nan2011}, we will have access to precisions which have never been explored. 

The improvement of the resolution of pulsar timing techniques has already made it possible to have information about the first seconds of a glitch \citep{PalfreymanVela} allowing, also, to fit the profile of the rise with simple theoretical models \citep{ashton2019rotational,pizzochero2019core} and constrain the glitch rise-time. In particular, the observation of the largest glitch of the Crab pulsar \citep{shaw+2018} has also allowed to put constraints on the mutual friction parameters \citep{Haskell2018Crab}. 
Furthermore, the coupling between the superfluid and the normal fluid plays a key role when extracting nuclear physics parameters related to the equation of state from measurements of the activity of a pulsar \citep{NewtonBergerHaskell2015}. Precise measurements of the activity can also be used  to constrain the mass of a pulsar \citep{Ho_2015,pizzochero+2017}. 
Thus, it is becoming increasingly important to have models which are able to produce quantitative predictions for the coupling time-scale between the normal and the superfluid component.  

General Relativity plays a fundamental role in neutron star theory as it is not possible to obtain realistic predictions for masses and radii in a merely Newtonian context \citep{shapiroteukolsky}. 
This implies that any quantitative study, whose results depend on the structure of the neutron star, must be performed considering the density profile to be a solution of the Tolman-Oppenheimer-Volkoff (TOV) equations, making non-relativistic models  inconsistent. Furthermore, \cite{Rothen1981} has shown, for a simplified constant density neutron star model, that the spacetime curvature can affect significantly the geometry of the vortex lines and this can have consequences on the estimates of the mutual friction. In addition, the difference of speed between the clocks on the Earth and the clocks in the neutron star, due to the gravitational redshift phenomenon, influences every dynamical time-scale we observe. 
It is clear, then, that to make quantitative predictions of the glitch rise-time, it is necessary to have an estimate of the relativistic effects. An initial study was performed by \cite{sourie_glitch2017} for a system composed of two rigid components, and significant quantitative differences (a correction of the order of $40 \%$) were found with respect to Newtonian models for selected equations of state.

Here, our final objective is to construct a consistent general relativistic model for the mutual friction coupling in neutron stars, that can be applied to pulsar glitches, r-mode damping \citep{Haskell2015R-modes}, asteroseismology and to the study of gravitational wave emission \citep{Glampedakis2018survey}. 
Thus, with the aim of quantifying the role of General Relativity, in this paper we will derive directly from Carter's multifluid formalism a local simple formula for the relativistic corrections to the coupling time. To achieve this goal we will also extend the vortex-tracing technique developed by \cite{Rothen1981} to make it applicable to the computation of the vortex density. Finally, using the conservation of the angular momentum, we will find a formula for the correction to the glitch rise-times.   
A key ingredient which makes this formula a universal relation (namely, the correction factor is independent on the choice of the equation of state) is the assumption that the angular momentum reservoir is located in a thin shell near the surface encompassing the crust and possibly parts of the outer core \citep{Ho_2015}. 
This puts our results in contrast to \cite{sourie_glitch2017}, where it is assumed that the reservoir is extended over the whole core.

In the derivation of the formula we follow a sequence of five steps which cover the different aspects of the problem.   

\textit{Section 2}: We review the basic ideas of Carter's hydrodynamic  formalism and of the prescription for vortex-mediated mutual friction of \cite{langlois98}, recasting the equations in a form which is convenient for our purposes.

\textit{Section 3}: We study the general mathematical properties of the macroscopic vorticity of a circularly rotating non-turbulent superfluid in a stationary axisymmetric system, deriving simple techniques to compute the vortex shape and density.

\textit{Section 4}: The techniques developed in the previous section are applied to the case of the neutron superfluid in a star. We derive analytic formulas for the vortex density and shape in the slow rotation approximation, and compare our results with \cite{Rothen1981}. The approximations proposed in \cite{Ravenhall_Pethick1994} are used to simplify the expressions, unveiling their physical interpretation.

\textit{Section 5}: We employ the results of the previous sections to see how the rise-time of a glitch is modified by General Relativity. Using the universality relation proposed in \cite{Rezzolla_Universal_2016} for the moment of inertia, we show that the relativistic factor we obtain is a pure function of the compactness of the star, so it is independent from the equation of state.

\textit{Section 6}: We show that the universality of the relativistic correction is a direct cosequence of the assumption that the free superfluid is located in a thin shell near the surface. This is done by proving that the formula for the rise-time found by \cite{sourie_glitch2017} reduces to ours in this limit.\\

Throughout the paper we adopt the spacetime signature $ ( - , +, + , + ) $, choose units with the speed of light $c=1$ and Newton's constant $G=1$, use greek letters $\nu$, $\rho$, $\sigma$... for coordinate tensor indexes. The sign of the volume form is chosen according to the convention $\varepsilon_{0123} = \sqrt{-g}$.

%The symbol $\varepsilon$ refers to the volume form $\varepsilon_{\nu_1 ... \nu_p} = \sqrt{|g|} \, \text{perm} \{ \nu_1 ,...,\nu_p \}$

\section{Relativistic vorticity and mutual friction}\label{STS}

We briefly introduce the two-fluid formalism  to model the dynamics of superfluid neutron star interiors. The coupling between the two fluids is provided by  the vortex-mediated mutual friction, for which we adopt the prescription of \cite{langlois98}, rewriting it in a form which is convenient for our purposes. To make the physical interpretation of the mutual friction clear, in subsection \ref{superfluids} we review the geometric properties of the macroscopic vorticity in General Relativity.

\subsection{The two-fluid formalism}\label{sec:The two-fluid formalism}

A realistic hydrodynamic description of a neutron star should take into account the existence of four components: $n_p^\nu$, the normal four-current, $n_n^\nu$, the four-current of the superfluid neutrons, $s^\nu$, the entropy four-current, and an electromagnetic component \citep{haskellsedrakian2017,chamel_super}. 
It is useful to introduce the rest-frame density associated with each component, in particular
\begin{equation}
n_p=  \sqrt{- n_p^\nu  n_{p\nu}} \spc n_n=  \sqrt{- n_n^\nu  n_{n\nu}},
\end{equation} 
and the four-velocities 
\begin{equation}
u_p^\nu \, = \, n_p^\nu / n_p \spc u_n^\nu \, = \, n_n^\nu / n_n,
\end{equation}
which are clearly normalized to $-1$.

The conservation of the baryon number implies that 
\begin{equation}
\nabla_\nu (n_p^\nu + n_n^\nu )=0
\end{equation}
and the second law of thermodynamics requires $\nabla_\nu s^\nu \geq 0$. 
In the following we will assume, either because chemical equilibrium is reached, or because the hydrodynamic processes considered are faster than the time-scale of the reaction \citep{Termo}, that
\begin{equation}
\nabla_\nu n_p^\nu = -\nabla_\nu n_n^\nu =0 \, .
\end{equation}
The hydrodynamic description must be consistent with the Einstein equations,
\begin{equation}\label{einstein}
G^{\nu \rho} = 8\pi T^{\nu \rho}_{(tot)}, 
\end{equation}
where $G_{\nu \rho}$ is the Einstein tensor and $T_{(tot)}^{\nu \rho}$ is the total energy-momentum tensor accounting for the presence of all the four components. 
To recover a two-component model for a neutron star interior we split the energy-momentum tensor as
\begin{equation}\label{TOT}
T_{(tot)}^{\nu \rho} = T^{\nu \rho} + T_{(ext)}^{\nu \rho} \, ,
\end{equation} 
where $T^{\nu \rho}$ represents the fluid contribution obtained by using a two-fluid zero-temperature formalism of the kind employed by e.g. \cite{andersson_comer2000}. 
In this theory the equation of state is given in terms of a master function\footnote{
For an interpretation of the master function $ -\mathcal{E}$ as a thermodynamic potential see \cite{Termo}.} \citep{noto_rel, andersson2007review}
\begin{equation}
 -\mathcal{E} (n_p^2, n_n^2, n_{np}^2),
\end{equation}
with $n_{np}^2 = -n_n^\nu n_{p\nu}$, leading to the definition of the momenta per particle
\begin{equation}\label{pqwoeiru}
\mu_{\nu}^p = -\dfrac{\partial \mathcal{E}}{\partial n_p^\nu}  \spc  \mu_{\nu}^n = -\dfrac{\partial \mathcal{E}}{\partial n_n^\nu},
\end{equation}
of the generalised pressure
\begin{equation}
\Psi = -\mathcal{E} - n_p^\nu \mu_{\nu}^p - n_n^\nu \mu_{\nu}^n
\end{equation}
and of the fluid stress-energy tensor
\begin{equation}
T\indices{^\nu _\rho} = \Psi \delta\indices{^\nu _\rho} + n_p^\nu \mu_{\rho}^p + n_n^\nu \mu_{\rho}^n.
\end{equation}
The tensor $T_{(ext)}^{\nu \rho}$, on the other hand, contains all the contributions which are not considered in $T^{\nu \rho}$, such as the finite temperature corrections due to the presence of $s^\nu$, the elastic part of the stress tensor in the crust and the electromagnetic energy-momentum. These parts play a negligible role in \eqref{einstein}, but they are fundamental in the study of the dynamics of $n^\nu_p$ and $n_n^\nu$. 
Taking the four-divergence of \eqref{einstein} and \eqref{TOT}, 
we have
\begin{equation}\label{BbBbalance}
\nabla_\rho T\indices{^\rho _\nu} = f_{(ext)\nu} \, ,
\end{equation}
where  the external force density is defined as
\begin{equation}
f_{(ext)}^\nu := -\nabla_\rho T^{\nu \rho}_{(ext)} \, .
\end{equation}  
As a result, we have that a mixture of charged, superfluid and possibly solid components at finite temperature can be conveniently described in terms of a zero-temperature two-fluid model subject to the action of an external force. 

In glitch models of the kind pioneered by \cite{baym+1969}, the external force is expected to play two main roles \citep[see e.g. the discussion in ][]{antonellipizzochero2017}. 
Firstly, there is a contribution in $f^\nu_{(ext)}$ that is assumed to enforce the proton-electron fluid and the crustal lattice rigid rotation: this is implicitly incorporated into glitch models by imposing that $\Omega_p$, the angular velocity of the normal $p$ component, depends only on time. 
Secondly, it exerts a torque on the two-fluid system, which can be interpreted as the local contribution to the braking torque, which is included into the evolution equation for the total angular momentum of the neutron star.  

\subsection{The macroscopic vorticity of a superfluid}\label{superfluids}

At the mesoscopic scale, the momentum $\mu^n_\nu$ is related to the superfluid order parameter $\phi$ according to the Josephson relation \citep{Carter2006}
\begin{equation}\label{gringo}
\mu^n_\nu = \dfrac{k}{2\pi} \partial_\nu \phi \, ,
\end{equation}
where $k=\pi \hbar$ to account for the Cooper pairing mechanism (if the superfluid were a Boson fluid, then $k=2\pi \hbar$). 
On scales smaller than the inter-vortex separation, the relation \eqref{gringo} implies that the four-vorticity 
\begin{equation}
\varpi_{\nu \rho} := (d\mu^n)_{\nu \rho}
\end{equation}
must be concentrated into vortex filaments and zero elsewhere.
However, in an astrophysical context we are interested in the dynamics of macroscopic matter elements crossed by several quantised vortices. Hence, we must locally average the momentum, and the relative vorticity, over a portion of fluid. 
In this way, if the vortex filaments are arranged  in a tangled configuration (as is expected if quantum turbulence develops in neutron star interiors, \cite{greenstein70,andersson_turbulence}), then it is in general not possible to reconstruct the vortex line configuration starting from the knowledge of the macroscopic vorticity field. 

For simplicity, we assume that turbulence is absent, so that the quantized vortices in each local matter element are parallel to each other. Given this condition, the region of spacetime occupied by the superfluid can be foliated by two-dimensional worldsheets which follow the profile of the  vortex lines. 
\cite{Stachel1980} has shown that these spacetime foliations are completely described by a bivector field $\mathcal{S}^{\nu \rho}$ (normalized as $\mathcal{S}^{\nu \rho} \mathcal{S}_{\nu \rho} =-2 $) such that
\begin{equation}\label{buzz}
{\star}\mathcal{S}_{\nu \rho} \, \mathcal{S}^{\rho \sigma} = 0   
\end{equation} 
and 
\begin{equation}\label{buzz2}
{\star}\mathcal{S}_{\nu \rho} \, \partial_\sigma \mathcal{S}^{\rho \sigma} = 0 \, ,
\end{equation}
where the symbol ${\star}$ is the Hodge duality operator, which acts on a generic p-form $\Sigma$ as 
\begin{equation}\label{yogi}
{\star}\Sigma_{\nu_1 ... \nu_{4-p}} = 
\dfrac{1}{p!} \varepsilon^{ \lambda_1 ... \lambda_p}_{\phantom{\lambda_1 ... \lambda_p}\nu_1 ... \nu_{4-p}} 
\Sigma_{\lambda_1 ... \lambda_p} \, .
\end{equation}
Equation \eqref{buzz} is an algebraic degeneracy condition: it tells us that $\mathcal{S}^{\nu \rho}$, as seen as an antisymmetric $4\times 4$ matrix, must have rank 2. 
This implies that the bivector $\mathcal{S}^{\nu \rho}$ is simple, i.e. there are two vector fields, say $u_C$ and $v_C$, such that
\begin{equation}\label{bubu}
\mathcal{S}^{\nu \rho} \, = \, u_C^\nu \,  v_C^\rho - u_C^\rho \,  v_C^\nu .
\end{equation}
The factor of two in the normalization condition implies that we can impose $u_C$ and $v_C$ to be orthonormal, while the minus sign tells us that one of the two vectors, conventionally $u_C$, is timelike. 
This bivector represents the unit surface element of the wordsheet and the condition \eqref{buzz2} is the requirement that all these surface elements mesh together smoothly.

It is easy to verify that if we want the macroscopic four-vorticity $\varpi_{\nu \rho}$ to come from an array of vortices which have the shape given by $\mathcal{S}$, then it must be true that
\begin{equation}\label{bazinga}
\mathcal{S}^{\nu \rho} = - \dfrac{{\star}\varpi^{\nu \rho}}{\varpi}  \spc \text{with}  \spc  \varpi := \sqrt{\dfrac{\varpi_{\lambda \sigma} \varpi^{\lambda \sigma}}{2}}.
\end{equation}
Equation \eqref{bazinga} can be rewritten into the form
\begin{equation}\label{pluto}
\varpi_{\nu \rho} \, = \, \varpi \, {\star}\mathcal{S}_{\nu \rho} 
\, =  \, \varpi \,\varepsilon_{\nu \rho \sigma \lambda} \, u_C^\sigma \, v_C^\lambda \, ,
\end{equation}
so that the kernel of $\varpi_{\nu \rho}$ corresponds to the linear combinations of $u_C$ and $v_C$.
The above relation allows to rewrite  \eqref{buzz} as
\begin{equation}\label{gandalf}
\varpi_{\nu \rho} \, \mathcal{S}^{\rho \sigma} \, = \, 0 \, ,
\end{equation}
which naturally leads to define the two orthogonal projectors
\begin{equation}\label{frodo}
\paral\indices{^\nu _\rho}:= \mathcal{S}^{\nu \lambda} \mathcal{S}_{\lambda \rho}  \spc  {\perp}\indices{^\nu _\rho} := \dfrac{\varpi^{\nu \lambda}\varpi_{\rho \lambda}}{\varpi^2} \, .
\end{equation}
%
%They project the four-vectors into two orthogonal planes and satisfy the obvious conditions 
%\begin{equation}\label{merry}
%\paral^2 =\paral  \spc {\perp}^2 ={\perp}  \spc  \paral {\perp}= {\perp} \paral =0
%\end{equation}
% and 
%\begin{equation}\label{pipino}
%\paral + {\perp} = \delta.
%\end{equation} 
%$\paral$ projects any four-vector onto the subspace which is tangent to the worldsheet, while ${\perp}$ on its complementary orthogonal. 
%
In appendix \ref{misurabile} we show that, given an observer $\mathcal{O}$ with four-velocity $u_\mathcal{O}$, the vector
\begin{equation}\label{lapseudobig}
\varpi_{\mathcal{O}}^\nu =  \varpi \, \mathcal{S}^{\nu \rho} u_{\mathcal{O} \rho}
\end{equation}
can be interpreted as 
\begin{equation}\label{ilfulcrodeldiscorso}
\varpi_{\mathcal{O}}^\nu \, = \, k \, \mathfrak{N}_{\mathcal{O}} \, v_{\mathcal{O}}^\nu \, ,
\end{equation}
where $\mathfrak{N}_\mathcal{O}$ is the local surface density of vortices measured by $\mathcal{O}$ and $v_\mathcal{O}^\nu$ is the unit vector directed along the local vortex array, as seen by $\mathcal{O}$. 
Furthermore, the vector
\begin{equation}\label{uvo}
u_{{V\mathcal{O}}} := \dfrac{\paral u_{\mathcal{O}}}{\sqrt{-g(\paral u_{\mathcal{O}},\paral u_{\mathcal{O}})}}
\end{equation}
defines the four-velocity of the vortices in the frame of $\mathcal{O}$. This velocity is constructed in a way that the relative three-velocity between the lines and the observer $\mathcal{O}$,
\begin{equation}\label{cmvnfjigro}
w_{V{\mathcal{O}}} := \dfrac{u_{V{\mathcal{O}}}}{\Gamma_{V{\mathcal{O}}}} - u_{\mathcal{O}}  ,
\end{equation}
is orthogonal to the vortex lines. Here,  $\Gamma_{V\mathcal{O}} = -g(u_\mathcal{O},u_{V\mathcal{O}})$ is the Lorentz factor associated with the relative speed $\Delta_{V\mathcal{O}} = \sqrt{g(w_{V{\mathcal{O}}},w_{V{\mathcal{O}}})}$ between $u_\mathcal{O}$ and $u_{V\mathcal{O}}$. 
As it is explained in appendix \ref{misurabile}, this Lorentz factor accounts for the length contraction phenomenon in the definition of the vortex density $\mathfrak{N}_\mathcal{O}$ measured by the observer $\mathcal{O}$, namely
\begin{equation}\label{bilbo}
\mathfrak{N}_\mathcal{O} \, = \, \mathfrak{N} \, \Gamma_{V{\mathcal{O}}} \, ,
\end{equation}
where
\begin{equation}
\mathfrak{N}  = \sqrt{\dfrac{\varpi_{\lambda \sigma} \varpi^{\lambda \sigma}}{2 \, k^2}},
\end{equation}
can be interpreted as the rest-frame vortex density.

Despite having used a terminology related to the presence of quantised vortex lines in a superfluid, 
we end this subsection by remarking that the construction described so far can be applied to a more general class of  fluids. % \citep{lichnerowicz1955}. 
In all situations in which $\varpi_{\nu \rho}$ can be computed as the exterior derivative of $\mu^n_\nu$, it is possible to use the fact $(d\varpi)_{\nu \rho \sigma} =0$ to prove that \eqref{buzz} implies \eqref{buzz2}.
This means that a given field  $\varpi_{\nu \rho}$ defines a worldsheet foliation if and only if there exists a time-like vector field $u_C$ such that
\begin{equation}\label{saruman}
u_C^\rho \varpi_{\rho \nu} =0 \, .
\end{equation} 
Therefore, given an arbitrary fluid for which the above equation is satisfied, it is possible to replace the words \textit{vortex line} with \textit{macroscopic vorticity line} everywhere in this subsection.
This result is discussed in detail in appendix \ref{CDTFIGRSPTM} by taking advantage of the language of force-free magnetohydrodynamics, based on the fact that reading the momentum $\mu^n_\nu$ as a vector potential leads to interpret the quantities  introduced in equations \eqref{lapseudobig} and \eqref{cmvnfjigro} as the magnetic field and the drift velocity respectively.

\subsection{The mutual friction coupling}\label{sec:The mutual friction coupling}

A model for the vortex-mediated mutual friction is one of the most important elements in pulsar glitch modelling.
In this work we follow the mutual friction prescription of \cite{langlois98}, which we briefly 
rederive with a geometrical argument.
According to \cite{langlois98}, the tensorial quantity
\begin{equation}\label{cretaceo}
f_{M\nu} := - n_n^\rho \varpi^n_{\rho \nu}
\end{equation}
provides the relativistic generalization of the Magnus force density \citep[see also ][]{Carter_Prix_Magnus, Andersson_Mutual_Friction_2016}.
The minus sign is chosen in a way that $f_M/\mathfrak{N}$ can be interpreted as the force per unit length exerted by the superfluid on the normal matter inside the core of the vortex. The norm of $f_M$ is
\begin{equation}\label{normando}
|f_M| \,= \, \sqrt{f_{M\rho} f_{M}^\rho  } \, = \, n_n \, k \, \mathfrak{N} \, \Gamma_{Vn} \, \Delta_{Vn} \, ,
\end{equation}
in accordance  with what it is expected by considering that  $\mathfrak{N}_n = \mathfrak{N} \, \Gamma_{Vn}$ is the density of vortices measured in the frame defined by $u_n$, see equation \eqref{bilbo}. 
Regarding the direction of $f_{M }$, it is immediate to see that it is orthogonal to $u_C$, $v_C$ and $u_n$.

In the presence of a normal component,  the vortex lines experience also a drag force per unit length \citep{Dbook1991}, that we will indicate as $\mathcal{F}_D$. To understand how this dissipative force can be modelled it is convenient to work in the frame 
%of the normal component. 
%According to \eqref{uvo}, in this frame the vortices move with 
defined by the four-velocity
\begin{equation}
u_{Vp} = \dfrac{\paral u_p}{\sqrt{-g(\paral u_p,\paral u_p)}} =: u_v \, .
\end{equation}
In this frame the vortices are at rest and the normal component moves orthogonally to the vortex lines, see the previous section. The vector $u_{Vp}$ can be interpreted as the proper four-velocity of the vortices, so, for notational convenience, it will be referred to as $u_v$, in full analogy with the notation used for $u_n$ and $u_p$. 

Since the  drag force on vortices has to balance the Magnus force (because the vortex lines have no inertia), it must be orthogonal to the vortex worldsheet. Assuming a viscous drag force per unit length $\mathcal{F}_D$ which is proportional to the three-velocity of the normal component in the frame of $u_v$, this results in
\begin{equation}\label{rtbjbnrtonbortlg}
\mathcal{F}_D = \alpha  \, \Gamma_{vp}^{-1} \,  {\perp} u_p  \,  ,
\end{equation}
where $\alpha$ is a coefficient that sets the strength of the microscopic dissipative interaction between a vortex core and the constituents of the normal component. 
%Note that, if $\alpha \neq 0$, then $\mathcal{F}_D =0$ if and only if $u_v=u_p$, as one would expect.
 
The averaged force per unit volume $f_D$ is found by multiplying the force exerted on a vortex by the number $\mathfrak{N}_p$ of vortices per unit area in the frame defined by  $u_p$,  
\begin{equation}
f_{D\nu} \,=\, \mathfrak{N}_p \, \mathcal{F}_{D\nu} 
\,=\, \alpha \,\mathfrak{N} \,{\perp}\indices{^\rho _\nu}u_{p\rho} \, ,
\end{equation}
see equation \eqref{bilbo}.
Hence, it is possible to write the force balance
\begin{equation}
f_M + f_D =0 \, ,
\end{equation}
which, defined the dimensionless factor
\begin{equation}
\mathcal{R} \, := \, \alpha \,  / ( k \, n_n ) \, ,
\end{equation}
takes the form
\begin{equation}\label{mutualfriction}
u_n^\rho \, \varpi^n_{\rho \nu} \, = \, \mathcal{R} \, k \, \mathfrak{N} \, {\perp}_{\nu \rho}  u_p^\rho,
\end{equation}
which is the one proposed in \cite{langlois98}. 
This equation, despite being clear from the geometrical point of view, is not written in the best form for practical purposes. 
By using the two orthonormal generators of $\mathcal{S}$, i.e. $u_v$ and $v_p$ ,
equation  \eqref{mutualfriction} can be cast as
\begin{equation}\label{mutualfriction2}
-\varepsilon_{\nu \mu \rho \sigma} u_n^\mu u_v^\rho v_p^\sigma = \mathcal{R}(u_{p\nu} -\Gamma_{vp}u_{v\nu}).
\end{equation}
In appendix \ref{sec:Reduction of the mutual friction} we show how to remove the dependence on $u_v$, so that the mutual friction is expressed in terms of the relative velocity between the $n$ and $p$ components. This procedure gives 
\begin{equation}\label{laforte}
-u_n^\rho \varpi^n_{\rho \nu} = \dfrac{\mathcal{R}k\mathfrak{N}}{1+\mathcal{R}^2} \bigg[  \mathcal{R} u_n^\lambda v_p^\sigma \varepsilon_{\lambda \sigma \nu \rho} w_{np}^\rho + \hat{\perp}_{\nu \rho}w_{np}^\rho  \bigg],
\end{equation}
for the non-relativistic $\Delta_{np} \ll 1$ limit of the Magnus force \eqref{cretaceo},
where  
\begin{equation}
w_{np} = \dfrac{u_n}{\Gamma_{np}} - u_p \approx u_n - u_p
\end{equation}
and $\hat{\perp}\indices{^\nu _\rho}$ is the projector orthogonal to the plane generated by $u_n$ and $v_p$. 
For small velocity lags between the superfluid and normal components ($\Delta_{np} \ll 1$), equation \eqref{mutualfriction} coincides with the one of \cite{Andersson_Mutual_Friction_2016}. 
In fact, equation \eqref{laforte} can be obtained also as the limit for non-relativistic relative speeds of equation (67) in \cite{Andersson_Mutual_Friction_2016}.

\section{Stationarity, axial symmetry and circularity condition}\label{saacc}

%The first step to a complete understanding of the relativistic effects on a superfluid system consists of analysing how they modify the density and the profile of the quantised vortex lines.
 
In this section we present the general properties of the macroscopic vorticity of a non-turbulent superfluid (introduced in section \ref{superfluids}) under the assumptions of axial symmetry, stationarity and circular motion. 
Given these three  assumptions, the following analysis is valid for a general fluid. Hence, only in this section we drop the label $n$ of the four-momentum $\mu^n$, since the superfluid species can be arbitrary. 
We specialise our results to a neutron star context in section \ref{corotante}.

%To achieve this aim we will study the four-vorticity in an arbitrary instant of the inter-glitch phase, in which the system can be conveniently approximated as stationary.

\subsection{Motion in a circular spacetime}\label{kerrmetric}

%In the following we will study processes in which the variation of the metric with time can be neglected, see \cite{langlois98}. 
Before studying the properties of the macroscopic vorticity we have to introduce the general form for the four-velocity of matter elements. In a stationary and axially symmetric spacetime the metric is invariant under transformations generated by a time-like Killing vector field $\xi$ and a space-like Killing vector field $h$ with closed orbits. 
The fields $\xi$ and   $h$ can always be set in a way that they commute, so it is possible to choose a chart such that \citep{Carter_Commutation}
\begin{equation}
\xi = \partial_t  \spc  h=\partial_\varphi \, .
\end{equation}
We also assume that the stationarity and axial symmetry properties are shared by the hydrodynamical quantities:  given a generic tensor $q$, we impose that $\mathcal{L}_\xi q = \mathcal{L}_h q=0$, where $\mathcal{L}$ is the Lie derivative \citep{gourRS}.

The final assumption is the  circularity condition, according to which  the currents of the chemical species in every point of the spacetime are  linear combinations of $\xi$ and $h$. Since  the corrections to the metric due to   $T_{(ext)}^{\nu \rho}$ are negligible, the  circularity condition implies that the two vectors $T\indices{^\nu _\rho} \xi^\rho$ and  $ T\indices{^\nu _\rho}h^\rho$ are linear combinations of $\xi$ and $h$ themselves \citep{andersson_comer2000}.

Given the above assumptions, there is always a collection of charts $(t,\varphi, x, y)$ such that the metric takes the form  \citep{Hartle_Sharp67}
\begin{equation}\label{astratto}
g_{\nu \rho} =
\begin{bmatrix}
   -N^2 + \omega^2 \rho^2 & -\omega \rho^2 & 0 & 0  \\
    -\omega \rho^2 & \rho^2 & 0 & 0  \\
   0 & 0  & g_x^2 & 0  \\
    0 & 0 & 0 & g_y^2
\end{bmatrix} \, .
\end{equation}
The lapse function $N$  contains information about the gravitational redshift, while the frame dragging $\omega$ is related to the Lense-Thirring effect. 
Furthermore, the physical quantities are functions of $x$ and $y$ only. 
%The square root of the determinant of this metric is
%\begin{equation}
%\sqrt{-g} = \rho N g_x g_y   
%\end{equation}
%and the inverse metric has the form
%\begin{equation}
%g^{\nu \rho} =
%\begin{bmatrix}
%  -N^{-2}  & -\omega N^{-2} & 0 & 0  \\
%   -\omega N^{-2} & \rho^{-2} - \omega^2 N^{-2} & 0 & 0  \\
%   0 & 0  & g_x^{-2} & 0  \\
%    0 & 0 & 0 & g_y^{-2}
%\end{bmatrix}.
%\end{equation}

The Zero Angular Momentum Observer (ZAMO) at a point is defined through the four-velocity
\begin{equation}
u_Z := N^{-1} (\partial_t + \omega \partial_\varphi) \, ,
\end{equation}
which is constructed in a way such that its dual is
\begin{equation}\label{perch lui}
u_Z^\flat  \, =  \,  -N dt  \, .
\end{equation}
In this way, the ZAMO is an Eulerian observer, namely an observer whose local set of simultaneous 
events is tangent to the surfaces $t=const$  \citep{rezzolla_book, Gourgoulhon3+1}.
In such a spacetime, a general four-velocity field of a matter element takes the form
\begin{equation}\label{circularfour}
u_{\mathcal{O}} 
\, = \, 
{N}^{-1} \, {\Gamma_{\mathcal{O}Z}} \, (\partial_t  + \Omega_\mathcal{O} \partial_\varphi) \, , 
\end{equation}
with
\begin{equation}\label{quarantetre}
\Gamma_{\mathcal{O}Z}^{-1}  \, = \, \sqrt{1-\Delta_{\mathcal{O}Z}^2}  
\spc 
\Delta_{\mathcal{O}Z} \, = \, \rho N^{-1} (\Omega_\mathcal{O} -\omega) \, .
\end{equation}
Here $\Delta_{\mathcal{O}Z}$ is the speed (with sign) of an observer moving with four-velocity $u_\mathcal{O}$,  measured in the frame of the ZAMO. 
In the following it will be useful to use the dual of $u_\mathcal{O}$, whose expression is 
\begin{equation}\label{ziopaperone}
u_\mathcal{O}^\flat
=
-\Gamma_{\mathcal{O}Z} (N+ \omega \rho \, \Delta_{ \mathcal{O} Z }) dt 
+ 
\Gamma_{\mathcal{O}Z} \Delta_{ \mathcal{O} Z } \, \rho \, d\varphi \, .
\end{equation}
As a result of the circularity condition, the four-velocity of each species can be written in the  form \eqref{circularfour} and its dual in the form \eqref{ziopaperone}.

%Now that the formalism has been set up, we can analyse the properties of the four-vorticity of a superfluid in a neutron star. The reference hydrodynamical system is the conventional two-fluid zero temperature model, in which we consider two number four-currents $n_n^\nu$ and $n_p^\nu$, where the species $n$ is the neutron superfluid  and the species $p$ is the proton-electron fluid (see il nostro). The momenta per particle of the two species are $\mu_{n\nu}$ and $\mu_{p\nu}$. Since in this section we have the purpose of analysing the geometrical structure of the vortices of a generic superfluid component in a stationary axisymmetric spacetime, without the need of specifying the details of the underlying hydrodynamical theory, we will deal with an abstract superfluid momentum $\mu_\nu$. This may clearly stand for the momentum of the $n$ component, but it may also denote the momentum of the $p$ one in the outer core.   

\subsection{Vorticity in a circular spacetime}\label{circularity}

Under the conditions imposed in the previous subsection, the momentum per particle of a species takes the form
\begin{equation}\label{ilmomentone}
\mu = \mu_t(x,y) dt + \mu_\varphi (x,y) d \varphi 
\end{equation}
and the corresponding four-vorticity is
\begin{equation}\label{immaturo}
\varpi_{\nu \rho} =
\begin{bmatrix}
  0  & 0 &  - \partial_x \mu_t &  -\partial_y \mu_t  \\
   0 & 0 & -\partial_x \mu_\varphi & -\partial_y \mu_\varphi  \\
    \partial_x \mu_t & \partial_x \mu_\varphi  & 0 & 0  \\
   \partial_y \mu_t & \partial_y \mu_\varphi & 0 & 0
\end{bmatrix}.
\end{equation} 
To unveil the underlying vortex structure, it is useful to consider  the function $\mathcal{N}(x,y)$ which counts the number of vortices enclosed in a loop $t,x,y=const$. 
The quantity $\mathcal{N}$ can be regarded as a rescaling of the azimuthal component of the momentum, as the Feynman-Onsager relation (see \eqref{motta} of appendix \ref{misurabile}) imposes that
\begin{equation}\label{quantizziamoinsieme}
\mu_\varphi = \dfrac{k\mathcal{N}}{2\pi} \, .
\end{equation}
We saw in subsection \ref{superfluids} that a four-vorticity  must have a non-trivial kernel, see equation \eqref{saruman}. This leads to the vanishing-determinant condition
\begin{equation}
\partial_x \mu_t \partial_y \mu_\varphi - \partial_x \mu_\varphi \partial_y \mu_t=0,
\end{equation} 
which can be alternatively written as
\begin{equation}\label{staomegagonza}
\dfrac{\partial_x \mu_t}{\partial_x \mu_\varphi} = \dfrac{\partial_y \mu_t}{\partial_y \mu_\varphi} =: -\Omega_C,
\end{equation}
where $\Omega_C$ is a function of $x$ and $y$. Therefore, employing both \eqref{quantizziamoinsieme} and \eqref{staomegagonza}, the vorticity in \eqref{immaturo} reads
\begin{equation}\label{maturo}
\varpi_{\nu \rho} = \dfrac{k}{2\pi}
\begin{bmatrix}
  0  & 0 &  \Omega_C  \partial_x \mathcal{N} &  \Omega_C  \partial_y \mathcal{N}  \\
   0 & 0 & -\partial_x \mathcal{N} & -\partial_y \mathcal{N}  \\
    -\Omega_C  \partial_x \mathcal{N} & \partial_x \mathcal{N}  & 0 & 0  \\
   -\Omega_C  \partial_y \mathcal{N} & \partial_y \mathcal{N} & 0 & 0
\end{bmatrix} \, .
\end{equation}
Moreover, defining the four-velocity
\begin{equation}\label{pulzella}
u_C := \dfrac{\Gamma_{CZ}}{N} (\partial_t + \Omega_C \partial_\varphi) \, ,
\end{equation}
we can verify that
\begin{equation}\label{strong}
u_C^\rho \varpi_{\rho \nu} =0 \, .
\end{equation}
It is possible to provide a simple physical interpretation of this result. 
First, there must exist a four-velocity $u_C$ which satisfies \eqref{strong} and an observer moving with this four-velocity will see the vortices at rest. Since we are considering a stationary configuration, the vortices cannot move towards the polar axis or back, because this would change $\mathcal{N}$. 
The only motion a vortex can undergo is a circular one around the axis, so that there must be a four-velocity satisfying equation \eqref{strong} with the form given in \eqref{pulzella}. 
The result is that, instead of dealing with $\mu_t$, we can directly consider $\Omega_C$, which describes the velocity of revolution of the vortices around the axis of the star. 

Now that we have ensured that the four-vorticity has a non-trivial kernel of dimension two, we can find a convenient basis for this space (the kernel defines the two-dimensional plane   tangent to the vortex worldsheet). Considering that $u_C$ satisfies equation \eqref{strong}, we need to compute only a second, linearly independent, basis vector. 
It is immediate to verify that
\begin{equation}\label{versorino}
v_C := \dfrac{\partial_y \mathcal{N} \partial_x - \partial_x \mathcal{N} \partial_y}{\sqrt{g_x^2 (\partial_y \mathcal{N})^2 + g_y^2 (\partial_x \mathcal{N})^2}},
\end{equation}
satisfies
\begin{equation}\label{nonmolliamomai}
v_C^\rho \varpi_{\rho \nu} =0   \spc g(v_C,v_C)=1  \spc  g(v_C,u_C)=0 ,
\end{equation}
so that we can express the projector in \eqref{frodo} as
\begin{equation}\label{ilparallelosemplice}
\paral\indices{^\nu _\rho} = -u_C^\nu u_{C\rho} + v_C^\nu v_{C\rho} \, .
\end{equation}
A final remark about the nature of $\Omega_C$ is needed: the condition \eqref{staomegagonza} implies
%It is equivalent to the couple of equations
%\begin{equation}\label{freaks}
%\partial_x \mu_t = -\Omega_C \partial_x \mu_\varphi  \spc  \partial_y \mu_t = -\Omega_C \partial_y \mu_\varphi
%\end{equation}
%We can derive the first with respect to $y$ and the second with respect to $x$. Then, if we subtract them, we get
\begin{equation}
\partial_x \mathcal{N} \, \partial_y \Omega_C - \partial_y \mathcal{N} \, \partial_x \Omega_C =0 \, ,
\end{equation}
which leads to  
\begin{equation}\label{scopertona!!!!}
v_C^\nu \, \partial_\nu \Omega_C \, = \, 0 \, .
\end{equation}
This means that the function $\Omega_C(x,y)$ cannot be arbitrarily chosen, but it must be conserved along the integral curves of $v_C$.

\subsection{Techniques to calculate the vortex structure}\label{profiliedensitingenerale}

Our purpose, now, is to provide simple techniques to visualise the global vortex structure, as well as useful formulas for its local properties. 

Consider a worldsheet of the spacetime foliation defined by $\varpi_{\nu \rho}$. Its intersection with a $t=const$ hypersurface is a space-like curve, which is the natural relativistic generalization of the concept of vortex line. We can parametrise it with the parameter $l$, chosen in a way that the tangent four-vector
\begin{equation}
\dfrac{d}{dl} := \dfrac{d x^\nu}{dl} \partial_\nu  
\end{equation}
is normalised to 1. Since $d/dl$ is tangent both to the worldsheet and to the constant time hypersurface, it must locally belong to the intersection between $span\{ u_C, v_C\}$ and $span \{ \partial_\varphi , \partial_x, \partial_y \}$. This, together with the normalization condition, implies that
\begin{equation}\label{mipiacequestolavoretto!}
\dfrac{d}{dl} = v_C \, ,
\end{equation}
if we choose the proper orientation. Therefore, the vortex lines are the integral curves of $v_C$, so that the vortices  lie in the plane $\varphi = const$. This fact is a consequence of the circularity condition and arises as a particular case of a more general result, which is presented at the end of this subsection. 
If we now apply \eqref{mipiacequestolavoretto!} to $\Omega_C$ and use \eqref{scopertona!!!!}, we see that 
 \begin{equation}\label{ilprimo}
 \dfrac{d \Omega_C}{d l} =0 \, ,
 \end{equation}
whose physical meaning  is clear: $\Omega_C$ at a point is the velocity of revolution of the unique vortex line passing exactly through that point. If $\Omega_C$  was varying along the same vortex line this would induce a deformation which would wrap  the vortex  around the rotation axis, breaking the stationarity of the system. Therefore, $\Omega_C$ must be constant along the vortex line, which is the physical interpretation of equation \eqref{scopertona!!!!}. This result is general and also applies to the angular velocity of the magnetic field lines in a stationary axisymmetric system, as discussed by \cite{Gralla_jacobson2014}.

As expected, the enclosed number of vortices $\mathcal{N}$ does not change along the profile of a vortex line. Formally, this is a consequence of \eqref{mipiacequestolavoretto!} and \eqref{versorino}, that can be combined to obtain
\begin{equation}\label{ilsecondo}
\dfrac{d \mathcal{N}}{d l} =0 .
\end{equation}
The immediate consequence  is that  the profile of the vortices coincides with the level curves of $\mathcal{N}(x,y)$.
In particular, the functions $\mathcal{N}$ and $\Omega_C$ are both constant on the vortex lines. This, together with the fact that $\mathcal{N}$ increases monotonically as we move far from the rotation axis, implies that we can parametrise the levels of $\Omega_C$ as $\Omega_C=\Omega_C (\mathcal{N})$, see also appendix \ref{deriv with forms} for a more detailed proof.

% In all the cases of interest, $\mathcal{N}$ is a function which increases as we depart from the axis, thus for any value of $\mathcal{N}$ there is only one curve in the intersection between the $t,\varphi =const$ half-plane and the region occupied by the superfluid such that $\mathcal{N}$ takes that value on it. On this curve, $\Omega_C$ will be forced to have a given uniform value, thus we can build the function $\Omega_C (\mathcal{N})$, which, for a given $\mathcal{N}$, finds the corresponding curve and returns the value that $\Omega_C$ assumes on it. 

Let us come back to equation \eqref{staomegagonza} and  rewrite it as
\begin{equation}
\partial_\nu \mu_t = -\dfrac{k \Omega_C }{2\pi} \partial_\nu \mathcal{N} \, .
\end{equation} 
Hence,  also $\mu_t$ can be expressed in terms of  $\mathcal{N}$ only and we have
\begin{equation}\label{chepwer!}
\mu_t (\mathcal{N}) =\mu_t(\mathcal{N} =0) -\dfrac{k}{2\pi} \int_0^{\mathcal{N}} \Omega_C (\mathcal{N}') d\mathcal{N}' \, .
\end{equation}
The overall constant $\mu_t(\mathcal{N} =0)$ can be fixed by considering that  $\mathcal{N}$ vanishes on the rotation axis, so that  $\mu_t$ is constant  there. Moreover, because of the circularity condition all the species comove on the axis and have a unique four-velocity $ u = N^{-1} \partial_t$ for all of them. 
As a consequence, the momentum per particle of a generic chemical species with chemical potential $\mu$ is $\mu_\nu =\mu u_\nu$ and  $\mu_t = -N\mu$ on the rotation axis, namely
\begin{equation}\label{urca}
\mu_t (\mathcal{N}=0) = -(N \mu)|_{axis} \, . 
\end{equation}
Let us now focus  on the local density of vortex lines. To visualize the densities obtained by taking a $t=const$ slice of the stellar interior we must consider the ZAMO introduced in  \eqref{perch lui}. The pseudovorticity vector associated to the ZAMO is
\begin{equation}
\varpi_Z^\nu \, = \, N \, ({\star}\varpi)^{\nu t} \, ,
\end{equation}
where the Hodge dual of the four-vorticity is
\begin{equation}\label{maturohodgiato}
{\star}\varpi^{\nu \rho} \! \! = \! \! \dfrac{k(2\pi\rho)^{-1}}{  N g_x g_y} \! \! \! 
\begin{bmatrix}
  0  & 0 &  -\partial_y \mathcal{N} &  \partial_x \mathcal{N}  \\
   0 & 0 & -\Omega_C \partial_y \mathcal{N} & \Omega_C \partial_x \mathcal{N}  \\
    \partial_y \mathcal{N} & \Omega_C \partial_y \mathcal{N}  & 0 & 0  \\
   -\partial_x \mathcal{N} & -\Omega_C \partial_x \mathcal{N} & 0 & 0
\end{bmatrix}  .
\end{equation}
%\begin{equation}\label{maturohodgiato}
%{\star}\varpi^{\nu \rho} \! = \! \dfrac{k(2\pi\rho)^{-1}}{  N g_x g_y} \!
%\begin{bmatrix}
%  0  & 0 &  - \mathcal{N}_{,y} &   \mathcal{N}_{,x}  \\
%   0 & 0 & -\Omega_C \, \mathcal{N}_{,y} & \Omega_C \, \mathcal{N}_{,x}  \\
%     \mathcal{N}_{,y} & \Omega_C \, \mathcal{N}_{,y}  & 0 & 0  \\
%   - \mathcal{N}_{,x} & -\Omega_C \, \mathcal{N}_{,x} & 0 & 0
%\end{bmatrix}  .
%\end{equation}
This explicit expression allows us to cast the  pseudovorticity   into the form 
\begin{equation}\label{oraepersempre}
\varpi_Z = \dfrac{k (2\pi\rho)^{-1}}{ g_x g_y} 
(\partial_y \mathcal{N} \, \partial_x - \partial_x \mathcal{N} \, \partial_y).
\end{equation}
A comparison with \eqref{ilfulcrodeldiscorso} and \eqref{versorino} gives
\begin{equation}\label{aaaaa}
v_Z = v_C
\end{equation}
and
\begin{equation}\label{zzzzz}
\mathfrak{N}_Z 
= 
 (2\pi \rho )^{-1} {\sqrt{ (g_x^{-1}\partial_x \mathcal{N})^2 +(g_y^{-1}\partial_y \mathcal{N})^2} } \, .
\end{equation}
This is the expression for the density of vortices in the frame of the ZAMO we were looking for. Note that \eqref{aaaaa} simply states that the local profile of the vortices seen by a ZAMO is tangent to $v_C$. 
%since the local set of simultaneous events in a ZAMO frame is tangent to a $t=const$ hypersurface, the vortex line (as  seen by a ZAMO) coincides with an integral curve of $v_C$, locally.

We can now easily calculate the rest-frame density of vortices. The definition \eqref{uvo} and the result \eqref{ilparallelosemplice} immediately give
\begin{equation}
u_{VZ} = u_C,
\end{equation}
meaning that the vortices move with four-velocity $u_C$ with respect to the ZAMO. 
The Lorentz factor associated with the relative motion is
\begin{equation}
\Gamma_{VZ} =\Gamma_{CZ} = \dfrac{1}{\sqrt{1 - \Delta_{CZ}^2}} 
 \spc 
  \Delta_{CZ} = \dfrac{(\Omega_C - \omega) \rho}{N},
\end{equation}
see \eqref{quarantetre}, and the vortex three-velocity in the frame of the ZAMO is
\begin{equation}
w_{VZ} = N^{-1} \, (\Omega_C - \omega) \, \partial_\varphi.
\end{equation}
Therefore, using \eqref{bilbo}, we have
\begin{equation}\label{zzzzzzzzzzzzzzzzzzzzzzz}
\mathfrak{N} = \Gamma_{CZ}^{-1} \, \mathfrak{N}_Z,
\end{equation}
where we recall that $\mathfrak{N}$ is the rest frame vortex density, while $\mathfrak{N}_Z$ is the vortex density in a ZAMO frame.
We can conclude that a ZAMO measures a vortex density which is increased with respect to the rest-frame one by a factor that encodes the relativistic length contraction effect due the vortex line motion around the rotation axis of the star.

%Once the rest-frame density of vortices is known, we can use \eqref{bazinga} and \eqref{giochiamocela} to write the bivector:
%\begin{equation}\label{maturohodgiatoridotto}
%\begin{split}
%\mathcal{S}^{\nu \rho} = 
%& \dfrac{\Gamma_{cz}}{ N \sqrt{ (g_y \partial_x \mathcal{N})^2 +(g_x \partial_y \mathcal{N})^2}} \cdot \\
%& \cdot \begin{bmatrix}
%  0  & 0 &  \partial_y \mathcal{N} &  -\partial_x \mathcal{N}  \\
%   0 & 0 & \Omega_C \partial_y \mathcal{N} & -\Omega_C \partial_x \mathcal{N}  \\
%    -\partial_y \mathcal{N} & -\Omega_C \partial_y \mathcal{N}  & 0 & 0  \\
%   \partial_x \mathcal{N} & \Omega_C \partial_x \mathcal{N} & 0 & 0
%\end{bmatrix},
%\end{split}
%\end{equation}
%which, remembering \eqref{pulzella} and \eqref{versorino}, is consistent with \eqref{bubu}.

We have shown that a vortex line lives on hypersurfaces $\varphi=const$. We conclude this section by showing how this property emerges exclusively from the circularity condition. At a given point, the intersection between the $t=const$ hypersurface and the vortex worldsheet trough the point is tangent to the pseudovorticity vector associated to the ZAMO. This is always true if the metric is Kerr-like, whether the circularity condition is satisfied or not. 
The vortex lines are thus  forced to lie on $\varphi =const$ surfaces if and only if  it is true that
\begin{equation}
\varpi_Z^\varphi = N ({\star}\varpi)^{\varphi t} =0 
\end{equation}  
everywhere. This is equivalent to saying that 
\begin{equation}
\varpi_{xy} =0 \, ,
\end{equation}
in complete analogy with the Newtonian case.

\section{Corotating case}\label{corotante}

We now specialise the analysis made in the previous section to the case of superfluid neutrons corotating with the proton-electron fluid in a neutron star.
%This is the simplest case and from the practical point of view it can be reduced to a single-fluid problem. 
We show that the vortex lines are almost straight in the quasi-Schwarzschild  coordinates when a realistic EOS is used, in contrast to the prediction of \cite{Rothen1981}, that is valid for an idealised star of constant density, in which the vorticity lines are found to have substantial curvature. 
 
The explanation of this fact is discussed by considering the approximations of \cite{Ravenhall_Pethick1994}, which are valid only for realistic EOSs and imply that the relativistic corrections to the profile of vortex lines cancel out. 
%On the contrary, if we use the unrealistic equation of state employed by \cite{Rothen1981}, these delicate cancellations do not occur. 

\subsection{Slow rotation approximation}\label{SRAPPX}

On the the dynamical time-scales we are interested in, the motion of the proton-electron fluid can be considered approximately rigid, so that $\Omega_p$ is a constant. Under the assumption of corotation we have that 
\begin{equation}
u_p =u_n= \dfrac{\Gamma_{pZ}}{N} (\partial_t + \Omega_p \partial_\varphi) 
\end{equation}
and by imposing chemical equilibrium we have
\begin{equation}\label{quantofacileora}
\mu_{\nu}^p = \mu_{\nu}^n = \mu u_{p\nu},
\end{equation}
where
\begin{equation}
\mu = \frac{\mathcal{E} + \Psi }{n_b},
\end{equation}
and the energy-momentum tensor reduces to that of a perfect fluid of baryons.

In the following we use  quasi-Schwarzschild  coordinates, obtained by requiring that $x=r$, $y=\theta$ are built in a such a way that the surfaces $t,r=const$ are conformal to a sphere, i.e. $g_\theta = \rho/\sin \theta$. 
The  quasi-Schwarzschild  coordinates have the advantage that, for slowly rotating neutron stars in which
\begin{equation}\label{gonzooooo}
\Gamma_{pZ} \approx 1 ,
\end{equation} 
one can make the simplifications
\begin{equation}\label{gonzissimooooo}
N\approx e^{\Phi(r)}   \spc    g_r \approx e^{\Lambda(r)}  \spc \rho \approx  r \sin \theta  \, .
\end{equation}
Here $\Phi$ and $\Lambda$ are solutions of the TOV equations, while the frame dragging is
\begin{equation}
\omega\,  = \,  \Omega_p \, \tilde{\omega}  \,  ,
\end{equation} 
where $\tilde{\omega}(r)$ can be found by using the approach of \cite{Hartle_slowly1}.
Therefore, the metric tensor has the form
\begin{equation}
ds^2 = -e^{2\Phi}dt^2 + e^{2\Lambda}dr^2 + r^2[d\theta^2 +  \sin^2 \theta(d\varphi - \omega dt)^2] 
\end{equation}
and the  generic circular four-velocity \eqref{circularfour} reads
\begin{equation}
u_\mathcal{O} = e^{-\Phi}(\partial_t + \Omega_\mathcal{O}\partial_\varphi)\, .
\end{equation}

\subsection{Vortex profile and density}\label{Kubo1}

We are now ready to explicitly calculate the profile of a vortex line in the simplest situation of a corotating body. 
In section \ref{circularity} we have seen that the function $\Omega_C$ is not completely arbitrary. 
However, it is easy to verify from \eqref{mutualfriction} that $\Omega_C = \Omega_p$ in the corotating case, so that the condition \eqref{scopertona!!!!} is fulfilled simply because $\Omega_p$ is uniform. 
This also allows to perform the integration in \eqref{chepwer!}: equations \eqref{quantizziamoinsieme} and \eqref{urca}, evaluated at the radius $r=R_D$ at which the neutrons start to drip out of the nuclei, give
\begin{equation}
\mu_{t}^n + \Omega_p \mu_{\varphi}^n = -m_n N_D \, ,
\end{equation} 
where $N_D:= N(r=R_D,\theta=0)$ and we have used the fact that at neutron drip $\mu=m_n$. 
We can make use of \eqref{quantofacileora} and \eqref{ziopaperone} to transform the above equation into a formula for $\mu$, 
\begin{equation}\label{chemicalgrosso}
\mu =\dfrac{m_n N_D}{N} \Gamma_{pZ}.
\end{equation}
Inserting these results into \eqref{quantizziamoinsieme} we finally arrive at 
\begin{equation}\label{ottantre}
\mathcal{N} = 2\pi k^{-1} m_n \dfrac{N_D \Gamma_{pZ}^2}{N^2} \rho^2 (\Omega_p - \omega) \, .
\end{equation}
%
%
%\subsection{Slow rotation approximation}\label{slowrotation}
%
%In order to plot the vortex structure, we need to use realistic expressions for the metric components which appear in \eqref{ottantre} and \eqref{zzzzz}. Since the two species are corotating, on a macroscopic scale they appear to be a single fluid with number four-current
%\begin{equation}
%n_b^\nu = n_p^\nu + n_n^\nu
%\end{equation}
%and enthalpy density
%\begin{equation}
%\dfrac{\mathcal{E} + \Psi}{n_b} = \mu,
%\end{equation}
%where $\mu$ is the equilibrium chemical potential of the two species presented in equation \eqref{quantofacileora}. Therefore, neglecting the electromagnetic contributions, we only have to solve the structure equations of a rotating relativistic star comprised of a single-species zero-temperature perfect fluid. Everything will be studied using the spherical coordinates: $x=r$, $y=\theta$. Thus we need to impose everywhere $g_x=E$ and $g_y=Fr$. The problem will be approached in slow rotation approximation, which implies that all the special relativistic effects of rotation are neglected,
%\begin{equation}
%\Gamma_{pZ} \approx 1,
%\end{equation} 
%the metric functions reduce to
%\begin{equation}
%N \approx e^{\Phi(r)} \spc  E \approx e^{\Lambda (r)} \spc  F \approx 1,
%\end{equation}
%where $\Phi$ and $\Lambda$ are solutions of the TOV equations, and the frame dragging is $\Omega_p$ multiplied by $\tilde{\omega}$, an adimensional factor, function only of $r$, found using the approach of \cite{Hartle_slowly1}. 
%
Within the slow rotation approximation this expression can be further simplified thanks to \eqref{gonzooooo} and \eqref{gonzissimooooo}, 
\begin{equation}\label{buzzlight}
\mathcal{N} = 2\pi k^{-1} m_n \Omega_p e^{\Phi_D - 2\Phi} (1 - \tilde{\omega}) r^2 \sin^2 \theta \, ,
\end{equation}
where $\Phi_D := \Phi (R_D)$. All the  relativistic corrections to $\mathcal{N}$ are contained  in the factor
\begin{equation}\label{lambdinopiccolo}
\lambda (r) :=  \left( 1-\tilde{\omega}(r) \right) \, e^{\Phi_D - 2\, \Phi(r)} \, ,
\end{equation}
meaning that
\begin{equation}\label{strangethings}
\mathcal{N} = \lambda (r) \, \mathcal{N}_{\textrm{Newt}} (\rho) \, ,
\end{equation}
where $\mathcal{N}_{\textrm{Newt}}$ is the vortex line density  for a non-relativistic superfluid system in uniform rotation \citep{FEYNMAN1955},
\begin{equation}
\mathcal{N}_{\textrm{Newt}} =  2\pi \, k^{-1} m_n \Omega_p r^2 \sin^2 \theta \, .
\end{equation}
It is possible to recognize  three different relativistic effects in the factor $\lambda$:
\begin{itemize}

\item \textit{Special relativistic dynamics}: since all the forms of energy contribute to the inertia, the relativistic momentum is $\mu \, u_\nu$ and not just $m_n \, u_\nu$. 
Thus, taking the slow rotation limit of \eqref{chemicalgrosso}, we expect a factor
\begin{equation}
\dfrac{\mu}{m_n} = e^{\Phi_D - \Phi}
\end{equation}
in the formula for $\mathcal{N}$. This factor, that grows as we move towards the center, has the effect to increase the momentum  (and, therefore, also the number of vortices) with respect to the Newtonian theory.

\item \textit{Gravitational dilation of times}: $\Omega_p$ represents the angular velocity of the neutron star as seen by a distant observer, which has a slow motion picture  of the internal dynamics. For an observer inside the star, everything is faster because of the gravitational dilation of time, so we expect the number of vortices $\mathcal{N}$ to be increased by a factor
\begin{equation}
\dfrac{dt}{d\tau} = e^{-\Phi}.
\end{equation}
This contribution behaves exactly as the previous one.

\item \textit{Frame dragging}: spacetime is distorted in a way that, from the point of view of a distant observer,  an Eulerian observer inside the star moves with angular velocity $\omega$. Hence, from Earth we see the superfluid rotating with angular velocity $\Omega_p$, but to compute the local properties of the vorticity field we have to subtract the apparent rotation $\omega$ of the Eulerian observer. This is the physical interpretation of the factor $( 1-\tilde{\omega})$, which has the effect of reducing the number of vortices (with respect to the Newtonian theory) and becomes smaller as we move towards the center, so that it partially cancels out the previous two contributions.
\end{itemize}
Now, recalling that the level curves of $\mathcal{N}$ are the profiles of the vortices, we immediately have that a vortex line passing trough the point $(r=r_{eq} , \theta=\pi/2)$ on the equatorial plane is defined by the implicit relation
\begin{equation}
r \, e^{-\Phi(r)} \sqrt{1 - \tilde{\omega}(r)}  \sin \theta 
= 
r_{eq} \, e^{-\Phi(r_{eq})} \sqrt{1 - \tilde{\omega}(r_{eq})} \, .
\end{equation}
This coincides with the early result of \cite{Rothen1981}, cf. equation (15) therein. This equation was obtained by \cite{Rothen1981} computing the integral curves of the pseudovorticity $\varpi_n^\nu$ in the frame of the superfluid itself for a single perfect-fluid model. This approach leads to the same formula we are presenting here because, from \eqref{maturohodgiato}, one can verify that  
\begin{equation}
\varpi_n^\nu=\Gamma_{pZ}^{-1}\varpi_Z^\nu,
\end{equation}
so $\varpi_n^\nu$ is proportional to $v_C^\nu$, see equation \eqref{aaaaa}.

We can, also, employ \eqref{zzzzz} to find the density of vortices in the frame of the ZAMO
\begin{equation}\label{chebbruttaragazzi!}
\mathfrak{N}_Z 
= 
\mathfrak{N}_{\textrm{Newt}} \, \lambda \sqrt{1+\sin^2 \theta 
\bigg[ e^{-2\Lambda} \bigg( 1 + \dfrac{r \partial_r \lambda}{2 \lambda}\bigg)^2 -1 \bigg]} \, ,
\end{equation}
where
\begin{equation}
\mathfrak{N}_{\textrm{Newt}} = 2 k^{-1} m_n \Omega_p 
\end{equation}
is the uniform density of vortices in the Newtonian limit.
For later purposes, it is worth mentioning here that \eqref{chebbruttaragazzi!}  reduces to 
\begin{equation}
\mathfrak{N}_Z = \mathfrak{N}_{\textrm{Newt}} \,   \lambda   
\end{equation}
on the polar axis $(\theta = 0)$, while  it reads
\begin{equation}\label{capitanfinduseilmare}
\mathfrak{N}_Z = \lambda \, \mathfrak{N}_{\textrm{Newt}}  \, e^{-\Lambda} \bigg( 1 + \dfrac{r \partial_r \lambda}{2 \lambda} \bigg) 
\end{equation}
on the equatorial plane $(\theta = \pi/2)$.

\subsection{Ravenhall and Pethick's approximation}\label{numericalresults}

\begin{figure}
\begin{center}
	\includegraphics[width=0.5\textwidth]{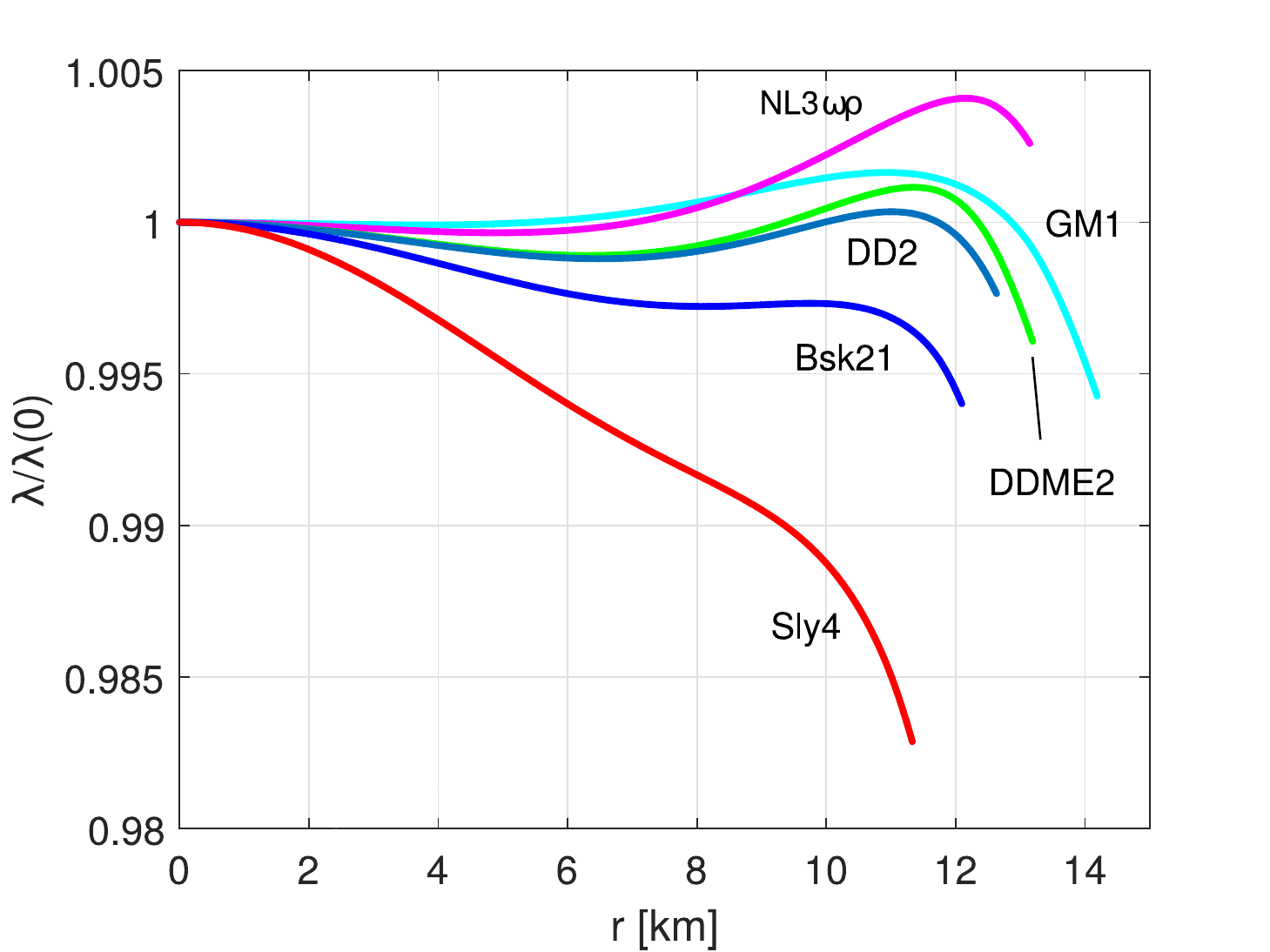}
	\includegraphics[width=0.5\textwidth]{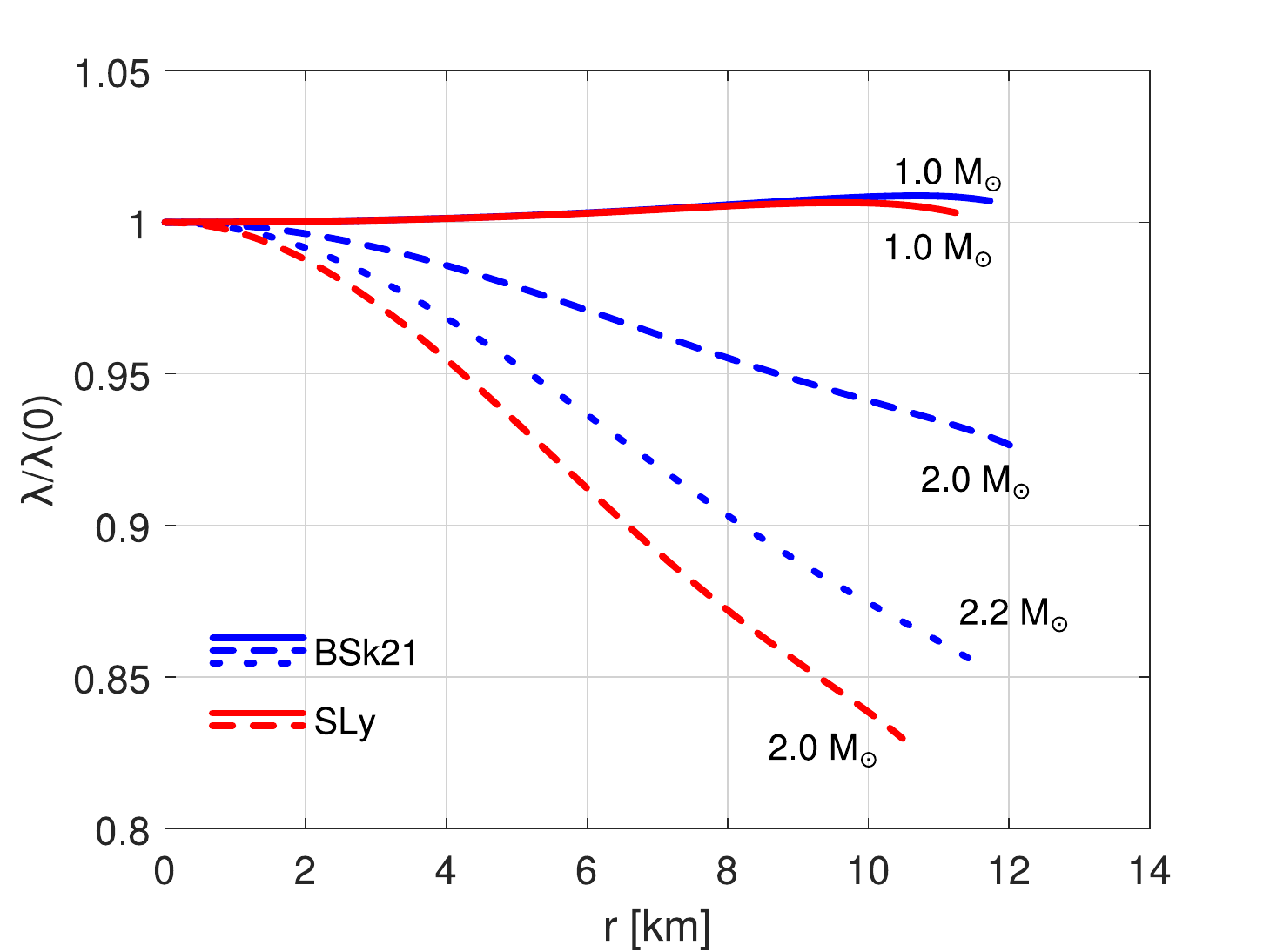}
	\caption{
	The relativistic factor $\lambda(r)$, normalised with respect to its value in the center $\lambda(0)$. In the upper panel we consider stars of $1.4 \, M_\odot$ for six different equations of state: NL3$\omega \rho$ (without hyperons), GM1 (without hyperons), DDME2 (with hyperons), DD2 (without hyperons), BSk21 and Sly4 \citep{Fortin2016,Fortin2017}. In the lower panel we show the profile of $\lambda/\lambda(0)$ for different (Komar) masses for two selected equations of state: $M = 1.0, \, 2.0, \, 2.2 \, M_\odot$ for BSk21 (\citealt{goriely+2010}) and $M = 1.0, \, 2.0\, M_\odot$ for SLy4 (\citealt{douchinhaensel2001}).
	} \label{fig:lambda}
	\end{center}
\end{figure}

The relativistic corrections to the vortex shape are encoded into the factor  $\lambda(r)$, see \eqref{strangethings}, which turns out to be approximately constant when realistic equations of state are used to integrate the TOV equations, as can be seen in figure \ref{fig:lambda}. 
It is interesting to show that this is a by-product of the validity of some approximations introduced by \cite{Ravenhall_Pethick1994}. 

\begin{figure}
\begin{center}
	\includegraphics[width=0.5\textwidth]{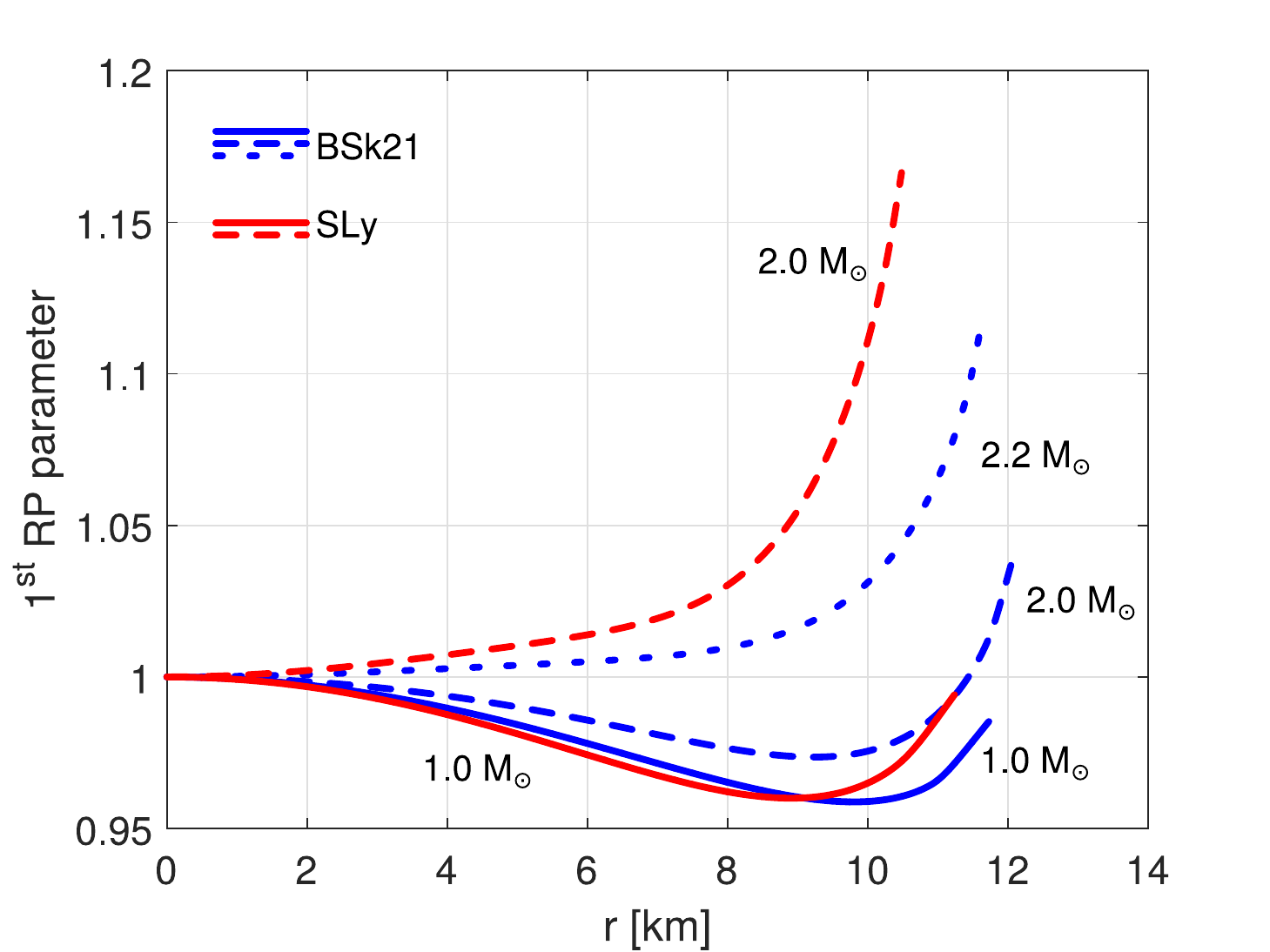}
	\caption{
	Plot of the first RP parameter, $e^{\Phi-\Lambda}$, normalised with its value in the center,  for different (Komar)  masses and equations of state: $M = 1.0, \, 2.0, \, 2.2 \, M_\odot$ for the BSk21 EOS (\citet{goriely+2010}, blue curves) and $M = 1.0, \, 2.0\, M_\odot$ for the SLy EOS (\citet{douchinhaensel2001}, red curves). }
	\label{fig:ephimenlambda}
	\end{center}
\end{figure}

Consider the quantity $e^{\Phi-\Lambda}$, which we call \textit{first Ravenhall and Pethick} (RP) \textit{parameter}. Using the TOV equations, it is immediate to see that
\begin{equation}\label{voldemort}
\dfrac{d}{dr}  \, e^{\Phi-\Lambda} 
= 
\dfrac{8}{3} \pi\, r\,  e^{\Phi + \Lambda}  
\bigg[ \braket{\mathcal{E}}_2 - \dfrac{3}{2} (\mathcal{E}-\Psi) \bigg],
\end{equation}
where $\braket{\mathcal{E}}_2$ is a volume average of the energy density,
\begin{equation}
\braket{\mathcal{E}}_2 = \dfrac{3}{r^3} \int_0^r  \, {r'\,}^2 \, \mathcal{E}(r') \, dr' .
\end{equation}
The right-hand side of equation \eqref{voldemort} goes to zero for $r=0$. However,   there is a competition between $\braket{\mathcal{E}}_2$ and $(3/2)(\mathcal{E}-\Psi)$  for $r>0$.
In the non-relativistic limit the pressure $\Psi$ is negligible with respect to $\mathcal{E}$, so that $3\mathcal{E}/2 > \braket{\mathcal{E}}_2$ in the region extending from the center to the radius at which $e^{\Phi-\Lambda}$ reaches a minimum. The minimum, however, is reached not far from the surface  for realistic equations of state, so that both $\braket{\mathcal{E}}_2$ and $3\mathcal{E}/2$ are small and $e^{\Phi-\Lambda}$ does not have the possibility to grow considerably. 
On the other hand, if the central pressure $\Psi(0)$ is comparable to the mass-energy density $\mathcal{E}(0)$, the inequality 
%$(3/2) (\mathcal{E}-\Psi) < \mathcal{E} $ may hold and, since the mass-energy density is always decreasing, this implies that 
\begin{equation}
\braket{\mathcal{E}}_2 \, > \,  3 (\mathcal{E}-\Psi) / 2
\end{equation}
may hold also for small values of $r$; in this case  $e^{\Phi-\Lambda}$ is an increasing function of the radial coordinate. 
Therefore,  we have two extremal situations in which the first RP parameter is always respectively lower and higher with respect to its central values. 
As we can see in figure \ref{fig:ephimenlambda}, neutron stars below their maximum  mass are exactly on the turning point between these two different behaviors. So we happen to be in the situation in which
\begin{equation}
e^{\Phi-\Lambda} \approx const.
\end{equation}  
Figure \ref{fig:ephimenlambda} also shows that the error which we commit with this approximation is of the order of $10\%$.

\begin{figure}
\begin{center}
	\includegraphics[width=0.5\textwidth]{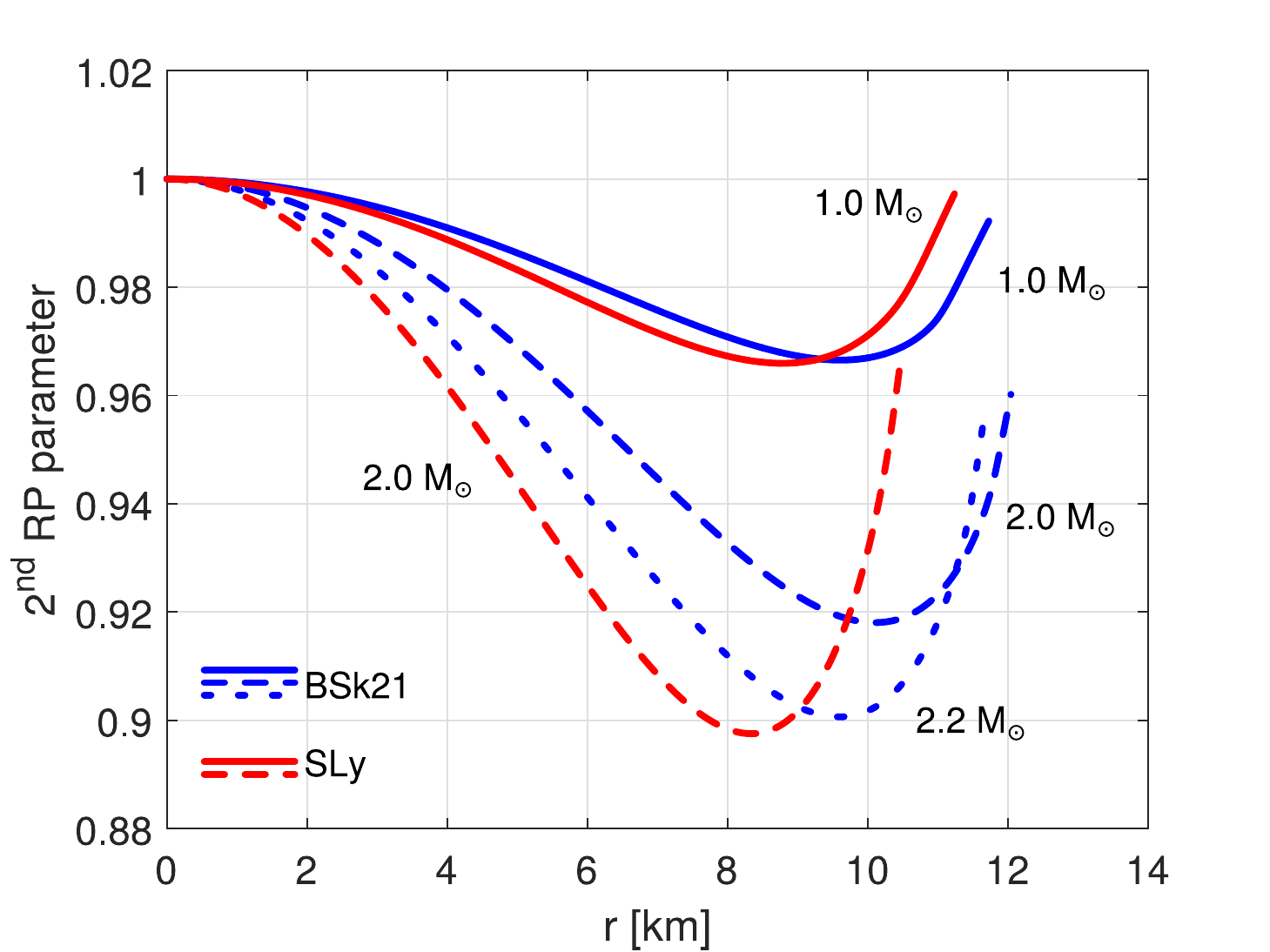}
	\caption{
	Plot of the second RP parameter, $e^{-\Phi-\Lambda}(1-\tilde{\omega})$, normalised with its value in the center,  for different (Komar) masses and equations of state: $M = 1.0, \, 2.0, \, 2.2 \, M_\odot$ for the BSk21 EOS (\citet{goriely+2010}, blue curves) and $M = 1.0, \, 2.0\, M_\odot$ for the SLy EOS (\citet{douchinhaensel2001}, red curves).
	}
	\label{fig:omegabarj}
	\end{center}
\end{figure}

Let us focus, now, on the quantity  $j(1-\tilde{\omega})$, which we call \textit{second Ravenhall and Pethick  parameter}, where $j:= e^{-\Phi-\Lambda}$. 
Given the equation for the frame dragging of \cite{Hartle_slowly1}, it is easy to see that
\begin{equation}\label{l'unicoanello}
\dfrac{d}{dr} \, [j(1-\tilde{\omega})] 
= 
(1-\tilde{\omega}) \dfrac{dj}{dr}  - \langle \,  
(1-\tilde{\omega}) \dfrac{dj}{dr} \,  \rangle_3 \, ,
\end{equation}
where
\begin{equation}
\langle \,  (1-\tilde{\omega}) \dfrac{dj}{dr} \,  \rangle_3
  = 
  \dfrac{4}{r^4} \int_0^r  (1-\tilde{\omega}(r')) \dfrac{dj}{dr'}\,  r'^3 dr' \, . 
\end{equation}
Now, the TOV equations allow to verify that  
\begin{equation}
(1-\tilde{\omega})\dfrac{dj}{dr} = -4\pi r (\mathcal{E}+\Psi)(1-\tilde{\omega}) e^{-\Phi + \Lambda}.
\end{equation}
For small $r$, both the terms in the right-hand side of equation \eqref{l'unicoanello} go to zero. As $r$ grows, $(1-\tilde{\omega})\partial_r j$ becomes negative, implying
\begin{equation}
(1-\tilde{\omega})\dfrac{dj}{dr} < \langle \,  (1-\tilde{\omega}) \dfrac{dj}{dr} \,  \rangle_3  \, .
\end{equation}
This average gives more importance to the contributions close to $r$, implying  that the difference $\partial_r j(1-\tilde{\omega}) - \braket{\partial_r j(1-\tilde{\omega})}_3$ is small. As we move towards the surface of the star, the fact that $\mathcal{E}+\Psi$ goes to zero becomes increasingly important and lowers the value of $\partial_r j (1-\tilde{\omega})$, until we reach a point in which $(1-\tilde{\omega}) \partial_r j= \braket{ (1-\tilde{\omega}) \partial_r j  }_3$. Here we have the minimum of $j(1-\tilde{\omega})$ after which it will start growing.
%Now it starts growing, but, again, for realistic equations of state this minimum is reached when we are not far from the surface, so this final growth is not sufficient to recover the difference that was accumulated in the first downhill. 
Thus, as can be seen in figure \ref{fig:omegabarj}, we can conclude that
\begin{equation}
e^{-\Phi-\Lambda}(1-\tilde{\omega}) \approx const 
\end{equation}
within an error of at most the $10\%$.

\begin{figure}
\begin{center}
	\includegraphics[width=0.5\textwidth]{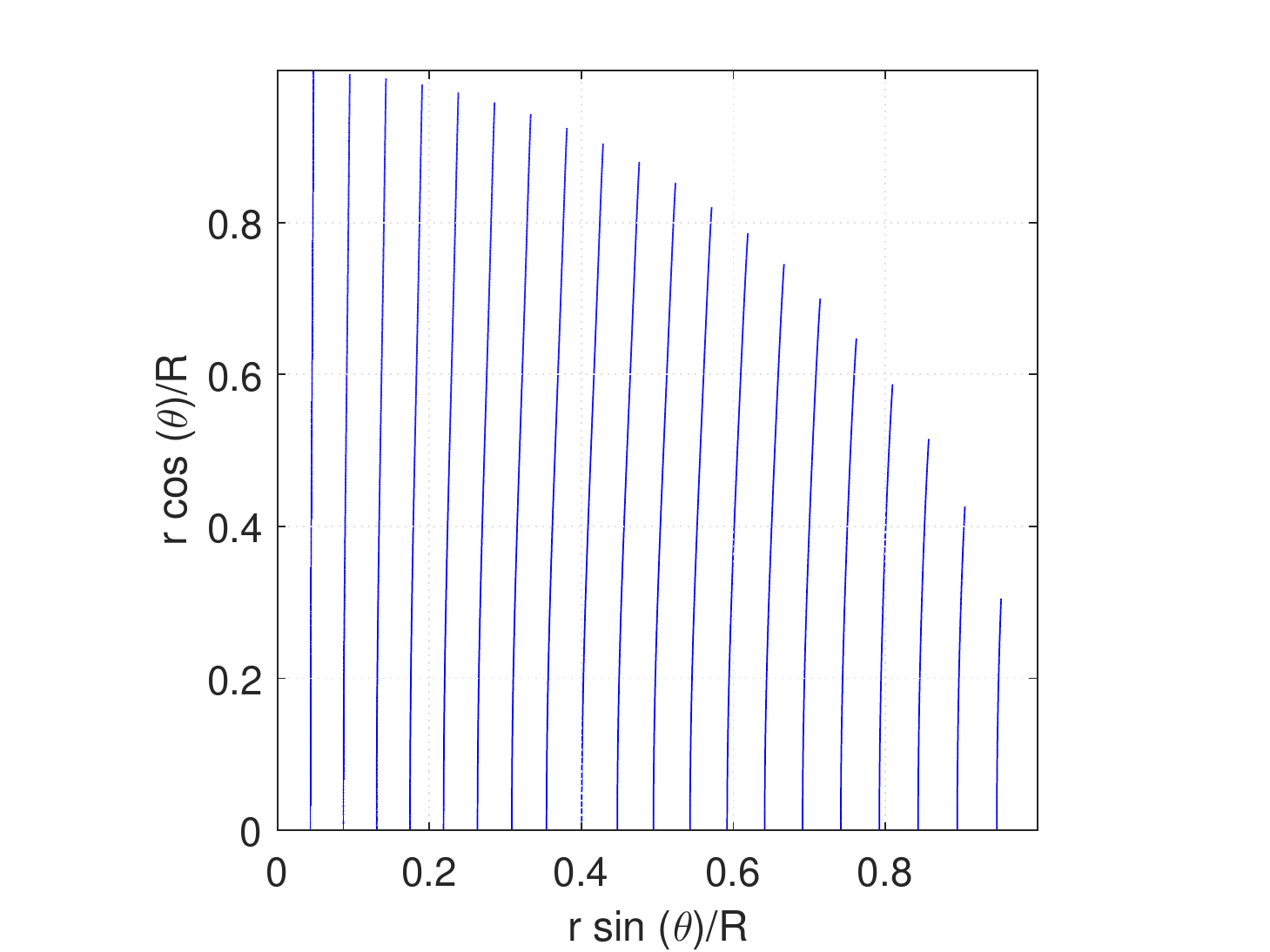}
	\caption{
		Vortex profile in the quasi-Schwarzschild coordinates for a neutron star of $2.0 \, M_\odot$ described by the SLy EOS. The spacing between vortices has been chosen in a way to facilitate the visualization and does not reflect the vortex density.}
	\label{fig:vortici}
	\end{center}
\end{figure}

Since $\lambda$, apart from an overall constant factor, is the ratio of the two RP parameters, then it is approximately constant as well (within the $2\%$ for masses below 1.4 $M_\odot$ and the $17\%$ for masses close to the maximum mass, see figure \ref{fig:lambda}). 
%
%There is, however, an interesting remark to make. As you can see comparing \eqref{fig:ephimenlambda} and \eqref{fig:omegabarj}, for low masses the profiles of $e^{\Phi-\Lambda}$ and $e^{-\Phi-\Lambda}(1-\tilde{\omega})$ are similar because both decrease for small $r$ until they reach a minimum. Therefore when we divide the two functions, their ratio tends to be a constant within an error of $2\%$. On the other hand, in the most relativistic stars the behaviour of the two factors is opposite, so errors on $e^{\Phi-\Lambda}$ and $e^{-\Phi-\Lambda}(1-\tilde{\omega})$ sum, giving a fluctuation of $\lambda$ which can reach $17\%$.
%
As a result, we do not present the plot of the vortex lines for the low mass cases because the relativistic effects are essentially invisible. We show only the profile for the most relativistic star considered here (i.e. a  2$M_\odot$ star with the SLy EOS, see figure \ref{fig:lambda}). The result is shown in figure \ref{fig:vortici}: the vortices are still essentially straight, as the gradients of $\mathcal{N}_{Newt}$ overwhelm the effect of the $\sim 20\%$ relativistic correction  due to  $\lambda$.

Finally, we remark that $\lambda \approx const$, which leads to almost straight vortex lines in this chart, is not the product of a more fundamental symmetry. Neutron stars supported by a realistic equation of state explore the particular range of parameters which guarantees this unexpected result. The use of unrealistic equations of state can lead to a highly deformed vortex structure. This is the case of  \cite{Rothen1981}, who employed the equation of state $\mathcal{E}=const$, which is pathological, especially close to the maximum mass where $\Psi(0)\longrightarrow \infty$.

%In \cite{Rothen1981}, however, the plots were obtained employing the equation of state $\mathcal{E}=const$, abandoning the delicate conditions for which the RP parameters are constant. To prove this it is enough to consider that the TOV equation
%\begin{equation}
%\dfrac{d \Psi}{d r} = -(\mathcal{E} + \Psi) \dfrac{d \Phi}{dr},
%\end{equation}
%combined with $\mathcal{E}=const$, immediately implies
%\begin{equation}
%e^\Phi = \dfrac{\mathcal{E}}{\mathcal{E} + \Psi} e^{\Phi (R)},
%\end{equation}
%where $R$ is the radius of the star. Thus if we take the first RP parameter, remembering that $e^{\Lambda(R)}=e^{-\Phi(R)}$ and $e^{\Lambda(0)}=1$, we find that
%\begin{equation}\label{tunztunztunz}
%\dfrac{e^{\Phi(0)-\Lambda(0)}}{e^{\Phi(R)-\Lambda(R)}} = \dfrac{\mathcal{E} e^{-\Phi (R)}}{\mathcal{E}+\Psi(0)}.
%\end{equation}
%When the mass approaches its maximum possible value, given by the condition
%\begin{equation}
%\dfrac{M}{R} = \dfrac{4}{9},
%\end{equation} 
%the pressure in the center of the neutron star diverges, sending the ratio \eqref{tunztunztunz} to zero. Therefore it is clear that when the star is considerably relativistic the RP approximation is not applicable for $\mathcal{E}=const$ and this results in the vortex profiles found by \cite{Rothen1981}. 

%\begin{figure}
%\begin{center}
%	\includegraphics[width=0.5\textwidth]{variousEOS2.pdf}
%	\caption{hjbhj
%		}
%	\label{fig:EOS}
%	\end{center}
%\end{figure}  

\begin{figure}
\begin{center}
	\includegraphics[width=0.5\textwidth]{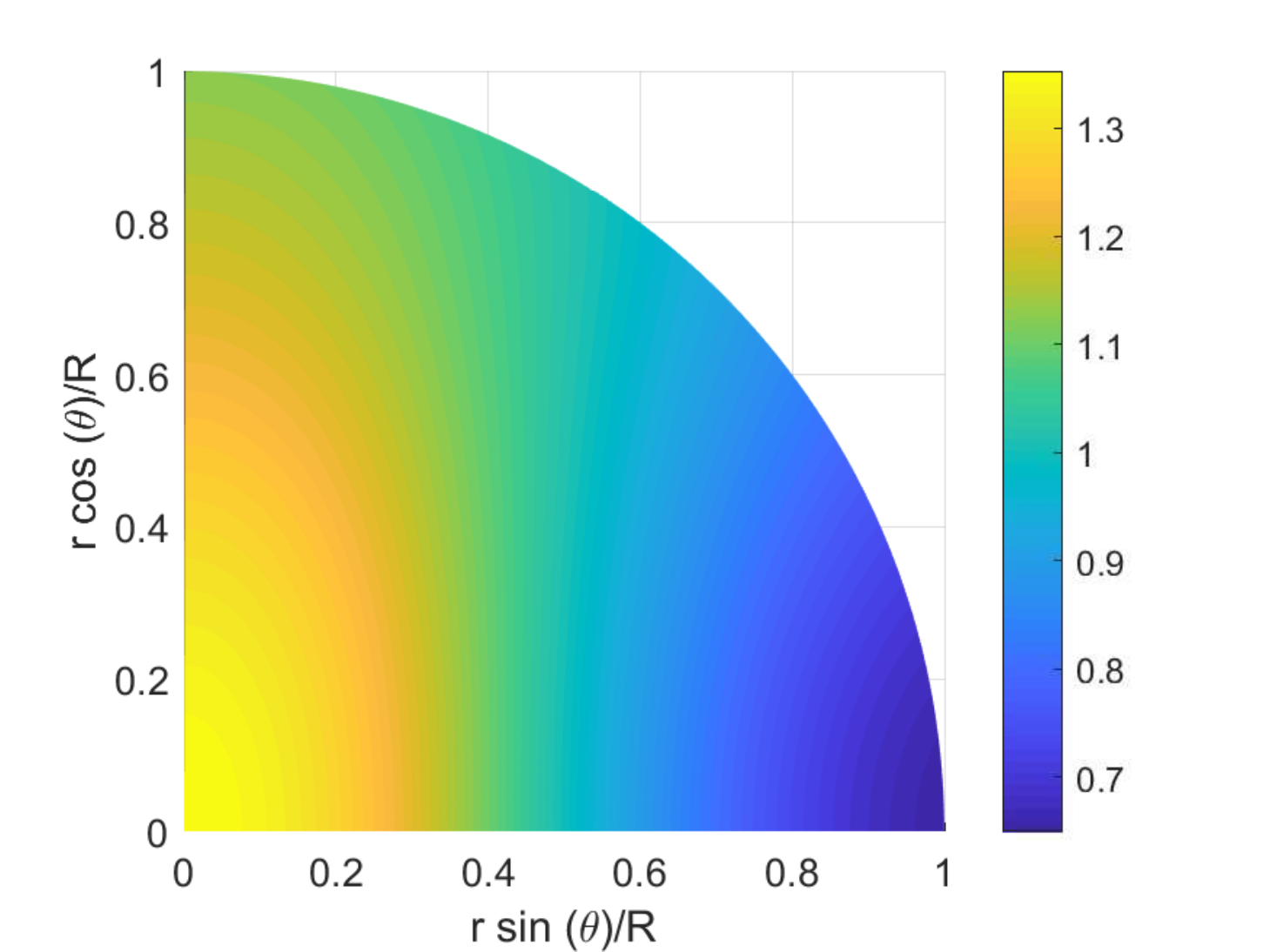}
	\caption{
		Map of $\mathfrak{N}_Z/\mathfrak{N}_{Newt}$ in the quasi-Schwarzschild coordinates, for a very compact neutron star of $2.0 \, M_\odot$ described by the SLy EOS.}
	\label{fig:colormapdensit}
	\end{center}
\end{figure}

\subsection{Approximate formula for the density of vortices}\label{approssimazionesulladensit}

We now focus on the vortex density given by equation \eqref{chebbruttaragazzi!}. In figure \ref{fig:colormapdensit} we plot the density of vortices for a $2 \, M_\odot$ neutron star governed by the Sly equation of state \citep{douchinhaensel2001}: it is higher close to  the axis of symmetry, while has a minimum at the equator. Let us consider this behaviour in more detail.

When Ravenhall and Pethick's approximation is applicable, we can put $\partial_r \lambda\approx 0$ in equation \eqref{chebbruttaragazzi!} to obtain
\begin{equation}\label{virgilio}
\mathfrak{N}_Z = \mathfrak{N}_{\textrm{Newt}} \lambda \sqrt{1+\sin^2 \theta ( -1 + e^{-2\Lambda} )}.
\end{equation}
There is a simple geometrical interpretation for this result: the fact that $\lambda$ is nearly constant means that the vortices are perfectly vertical and uniformly distributed in the chart (this is analogous to the Newtonian case, since the only correction is given by the overall rescaling factor $\lambda$). However, when we have to compute the density $\mathfrak{N}_Z$ measured by the ZAMO, we need to consider also the fact that the lengths are distorted and that the density must be computed orthogonally to the vortex lines.

Consider two different ZAMOs, one on the polar axis and the other on the equatorial plane. In the first case the induced metric on the surface orthogonal to the vortices is  
\begin{equation}
d\sigma^2 = r^2 d\theta^2 + r^2 \sin^2 \theta d\varphi^2,
\end{equation}
which coincides with its Newtonian limit. Therefore, $\lambda$ is the only relativistic correction and
\begin{equation}
\mathfrak{N}_Z = \mathfrak{N}_{\textrm{Newt}} \, \lambda.
\end{equation}
In the second case the surface element is
\begin{equation}
d\sigma^2 = e^{2\Lambda} dr^2 + r^2 d\varphi^2,
\end{equation}
so that the space is dilated in the radial direction by a factor $e^\Lambda$ with respect to the Newtonian limit. As a consequence, the measure of surface density becomes
\begin{equation}\label{mistermuscoloidraulicogel}
\mathfrak{N}_Z= \mathfrak{N}_{\textrm{Newt}} \, \lambda \, e^{-\Lambda}.
\end{equation}
% So let us consider a ZAMO located on the polar axis on a radius $r$. The vortices are locally parallel to the axis of the star, so the surface he is going to consider is an infinitesimal fragment of $t,r=const$. The induced metric on this sphere is
%\begin{equation}
%d\sigma^2 = r^2 d\theta^2 + r^2 \sin^2 \theta d\varphi^2,
%\end{equation}
%which coincides with its Newtonian limit. Therefore in the Newtonian and in the relativistic case the observers will agree on the value surface area. The number of vortices, however, will be rescaled by a factor $\lambda$, giving
%\begin{equation}
%\mathfrak{N} = \mathfrak{N}_{Newt}\lambda,
%\end{equation}
%which is coherent with \eqref{virgilio}. If, on the other hand, we consider a ZAMO who is located on the equatorial plane, then, considering that the lines are orthogonal to this plane, we need to take a fragment of this plane, whose induced metric is
%\begin{equation}
%d\sigma^2 = e^{2\Lambda} dr^2 + \rho^2 d\varphi^2.
%\end{equation}
%This tells us that the space in the radial direction is dilated in GR by a factor $e^\Lambda$, whose consequence is that the measure of surface density will become
%\begin{equation}\label{mistermuscoloidraulicogel}
%\mathfrak{N}= \mathfrak{N}_{Newt} \lambda e^{-\Lambda}.
%\end{equation}
%The factor $e^{-\Lambda}$ gets smaller as we approach the equator. 
Therefore, the square root factor in \eqref{virgilio} accounts for the fact that space is distorted: even if in the chart the vortices may be uniformly distributed (similarly to the Newtonian case), we still measure different local densities. The gravitational dilation of space in the radial direction is responsible for an alteration of the surface density, but only if the surface element has a radial extension, which explains the dependence on both $r$ and $\theta$ presented in equation \eqref{virgilio}.

\begin{figure}
\begin{center}
	\includegraphics[width=0.5\textwidth]{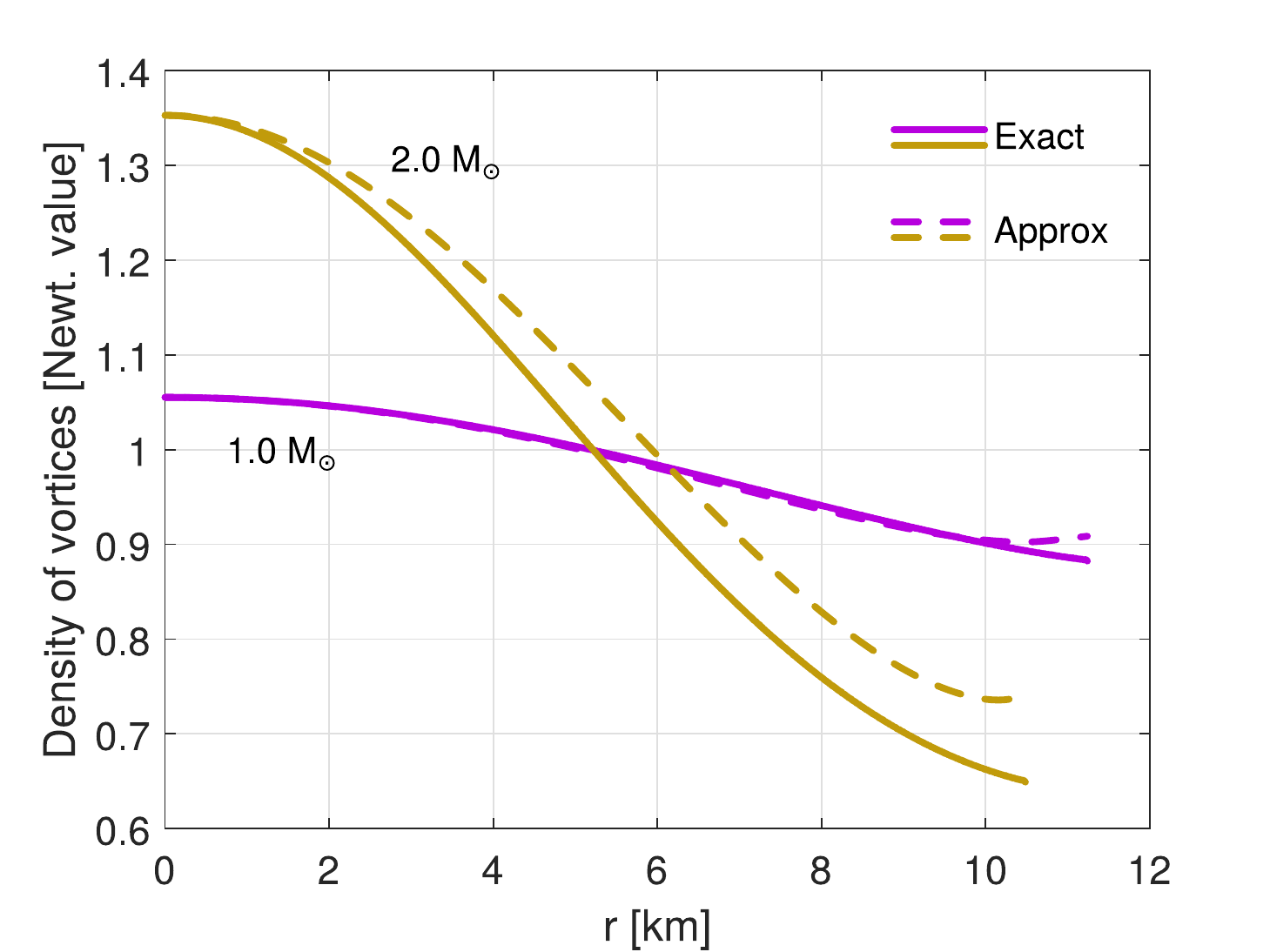}
	\caption{
		Plot of $\mathfrak{N}_Z/\mathfrak{N}_{Newt}$ as a function of $r$ along the equatorial plane ($\theta= \pi/2$), assuming the SLy equation of state.  The continuous lines represent the exact surface density given by equation \eqref{chebbruttaragazzi!}, the dashed ones represent the approximate result in equation \eqref{virgilio}.}
	\label{fig:densitequatoriali}
	\end{center}
\end{figure}

In figure \ref{fig:densitequatoriali} we show the profile of the density of vortices on the equatorial plane for different solar masses, assuming the SLy equation of state. In the same figure, we also compare the exact formula \eqref{capitanfinduseilmare} to the approximate one,  \eqref{mistermuscoloidraulicogel}. Clearly, the approximation is acceptable for low mass stellar configurations, but the induced error can be relevant for more massive stars. In all cases the approximation becomes better as we move towards the center, due to the fact that we are neglecting a term which is proportional to $r \partial_r \lambda$. 

\section{Relativistic correction to the coupling time-scale}\label{the end}

We are now ready to estimate the time-scale on which the mutual friction couples the superfluid and the normal components in a relativistic star. For a given velocity lag between the two components, we derive the time-scale on which the mutual friction couples them and re-establishes corotation. The time-scale we calculate is, therefore, also the one associated to the potentially observable spin-up phase in a pulsar glitch.

\subsection{Small lag approximation}\label{SMLapprox}

As a first step, we need to find the general expression for $\mu^n_\varphi$ in the presence of a velocity lag between the components.
Given that $\Omega_p$ is uniform and represents the angular velocity of the neutron star we see from the Earth, let us consider the presence of a lag $\Omega_{np}:=\Omega_n-\Omega_p$ such that
\begin{equation}\label{sszmall}
\dfrac{|\Omega_{np}|}{\Omega_p} \ll 1 ,
\end{equation}
so we can treat it as a perturbation. We also impose that the lag is different from zero only in a thin spherical shell (containing the crust), implying that the inertia of the part of superfluid which is not rotating with $\Omega_p$ is small compared to the rest of the star. 
This allows us to neglect the effect of this small lag on the metric.

The momentum per particle of the neutron superfluid can be written with the aid of an entrainment parameter $\epsilon_n$ \citep{prix2005} as
\begin{equation}\label{momentaaaaaaaAAAAAaaaaa}
\mu_{\nu}^n = \mu_n^{(C)} \bigg[(1-\epsilon_n)u_{n\nu} + \dfrac{\epsilon_n}{\Gamma_{np}} u_{p\nu} \bigg], 
\end{equation}
where $\mu_n^{(C)}$ is the \emph{comoving} chemical potential \citep{Termo}. 
%Hence, thanks to the slow rotation approximation we find
%\begin{equation}
%u_p^\nu \mu^n_\nu = \mu_n^{(C)}\bigg[ 1+\bigg(\dfrac{1}{2}-\epsilon_n\bigg)\Delta^2 \bigg].
%\end{equation}
%Furthermore, we assume that the lag is also so small that it does not change the thermodynamic state of the corotating system\footnote{The maximum value that $|\epsilon_n|$ is expected to assume in the crust is around 10 \citep{chamel2012}, so the approximation is correct provided that $|\epsilon_n| \Delta^2 \ll 1$, so $\Delta \ll 0.32$, which is always satisfied in a neutron star.}, namely
%\begin{equation}
%\Delta^2 \dfrac{\partial \mathcal{E}}{\partial \Delta^2} \ll \mathcal{E},
%\end{equation}
%which allows to make the approximation
%\begin{equation}
%u_p^\nu \mu^n_\nu \approx \mu_n^{(C)} \, .
%\end{equation}
%Therefore, the left-hand side of \eqref{focaccia} can be well approximated by means of $\mu_n^{(C)}$. The right-hand side can be simplified by using the slow rotation approximation, so that equation \eqref{focaccia} becomes
We have shown that, when the species corotate and are in chemical equilibrium, equation \eqref{chemicalgrosso} holds. In the slow rotation approximation this condition can be rewritten as
\begin{equation}\label{jJjJjJj}
\mu_n^{(C)} = m_n e^{\Phi_D - \Phi}.
\end{equation}
This equation remains true also in the presence of a lag, provided that
\begin{equation}
\bigg|\Delta^2 \dfrac{\partial \mu_n^{(C)}}{\partial \Delta^2} \bigg|_{n_p,n_n} \bigg| \ll \mu_n^{(C)}.
\end{equation}
It is easy to show that the above condition is equivalent to
\begin{equation}
|\epsilon_n| \Delta^2 \ll 1,
\end{equation}
which is verified in every realistic situation.\footnote{The maximum value that $|\epsilon_n|$ is expected to assume in the crust is around 10 \citep{chamel2012}, so the approximation is correct provided that $\Delta \ll 0.32$, which is always satisfied in a neutron star.}
Thus, we can plug equation \eqref{jJjJjJj} into equation \eqref{momentaaaaaaaAAAAAaaaaa} and use the slow-rotation approximation in \eqref{ziopaperone}, obtaining 
\begin{equation}\label{hhhhhhhh}
\mu^n_\varphi =  m_n e^{\Phi_D - 2 \Phi} \bigg[ (1-\tilde{\omega})\Omega_p + (1-\epsilon_n)\Omega_{np} \bigg] r^2 \sin^2 \theta.
\end{equation}
Clearly, when $\Omega_{np}=0$ the above relation becomes equation \eqref{buzzlight}. 

We remark that if the lag was different from zero for every $r$, and not only in a thin shell of small moment of inertia, the frame dragging would be modified by the lag and would have the form $\omega=\tilde{\omega}\Omega_p + \omega'$ where $\omega'=\omega'[\Omega_{np}]$ is a non-local linear functional of the lag. So in equation \eqref{hhhhhhhh} a term $-\omega'$ would appear inside the square brackets, which has the same order of $\Omega_{np}$ and therefore cannot be neglected. The consequences of this complication are discussed in section \ref{CompSour}.

\subsection{Local relaxation time}

We compute here the timescale with which the velocity lag in different rings  of constant $r$ and $\theta$ relax to corotation.

In the slow rotation approximation the relative velocity between the two components is
\begin{equation}
w_{np}= e^{-\Phi}\Omega_{np} \, \partial_\varphi.
\end{equation} 
Considering that $\hat{\perp}_{\varphi \varphi} \approx r^2 \sin^2 \theta$, the $\varphi$ component of equation \eqref{laforte} reads
\begin{equation}\label{parmuUU}
\partial_t \mu_{\varphi}^n = -\dfrac{\mathcal{R}k\mathfrak{N}_Z }{1+\mathcal{R}^2} \Omega_{np} r^2 \sin^2 \theta.
\end{equation}
Note that this equation coincides with equation (85) of \cite{langlois98}, neglecting the convection angular velocity $\Omega_+$ and using the slow rotation approximation. 
Recalling equation \eqref{hhhhhhhh}, we find that
\begin{equation}\label{sazu}
\partial_t \mu_{\varphi}^n =m_n e^{\Phi_D -2\Phi} [(1-\tilde{\omega})\dot{\Omega}_p +(1-\epsilon_n) \partial_t \Omega_{np}] r^2 \sin^2 \theta.
\end{equation}
Since the relaxation process we are considering is fast compared to the spin-down time-scale, the Komar angular momentum is approximately conserved. Moreover,  we are assuming that the part of superfluid which is not locked to the normal component is contained in a thin shell of small moment of inertia. Therefore, the angular momentum conservation implies
\begin{equation}
|\dot{\Omega}_p| \ll |\partial_t \Omega_{np}|,
\end{equation}
which allows us to neglect the term $(1-\tilde{\omega})\dot{\Omega}_p$ in  \eqref{sazu}. 
With the aid of this approximation, we can obtain from \eqref{parmuUU} a closed equation for the evolution of $\Omega_{np}$:
\begin{equation}
\partial_t \Omega_{np} = - \dfrac{\mathcal{R}k\mathfrak{N}_Z e^{2\Phi -\Phi_D}}{m_n(1+\mathcal{R}^2)(1-\epsilon_n)} \Omega_{np}.
\end{equation}
Integrating this equation we obtain that for each ring of constant $r$ and $\theta$ we have a different relaxation time-scale,
\begin{equation}
t_{R,\textrm{GR}}^{(ring)} = \dfrac{m_n (1+\mathcal{R}^2)(1-\epsilon_n)}{\mathcal{R}k \mathfrak{N}_Z e^{2\Phi - \Phi_D}}.
\end{equation}
Finally, isolating the Newtonian contribution
\begin{equation}
t_{R,\textrm{Newt}}^{(ring)} = \dfrac{m_n (1+\mathcal{R}^2)(1-\epsilon_n)}{\mathcal{R}k \mathfrak{N}_{\textrm{Newt}} },
\end{equation}
we find, using \eqref{virgilio} and \eqref{lambdinopiccolo},
\begin{equation}\label{dipendeontheta}
t_{R,\textrm{GR}}^{(ring)} (r,\theta)= \dfrac{t_{R,\textrm{Newt}}^{(ring)}(r)}{ (1-\tilde{\omega}(r)) \sqrt{1+\sin^2 \theta ( -1 + e^{-2\Lambda(r)} )}}.
\end{equation}
Hence, General Relativity introduces a dependence of the coupling time-scale on $\theta$ which is not present in the Newtonian limit. This is a direct consequence of the discussion in subsection \ref{approssimazionesulladensit}.

It is important to remark that our formula for the time-scale has been derived under the fundamental assumption of a mutual friction which is proportional to the modulus of the macroscopic vorticity, see subsection \ref{sec:The mutual friction coupling}. In a turbulent regime this assumption is likely to be violated and this might produce different contributions to the relativistic correction. To compute them, one can perform the same calculations we made in this subsection, modifying the right-hand side of \eqref{parmuUU} according to the alternative prescription for the mutual friction.

\subsection{Global relaxation time}\label{Kubo2}

Different rings reach corotation on a different time-scale, but from the Earth we can observe only the changes of $\Omega_p$. Its evolution can be obtained by using angular momentum conservation, so its behaviour will not be a simple exponential but a weighted average of exponentials. We expect the average to favour those rings which are located around $\theta=\pi/2$ (the equatorial plane), as they are those with the largest moment of inertia. Therefore, employing also the fact that the metric functions vary slowly in the thin shell and can, thus, be evaluated at $r=R_D$, we obtain
\begin{equation}\label{dododavecchio}
\dfrac{t_{R,\textrm{GR}}}{t_{R,\textrm{Newt}}} = \dfrac{e^{\Lambda_D}}{1-\tilde{\omega}_D},
\end{equation}
where $t_{R,\textrm{GR}}$ and $t_{R,\textrm{Newt}}$ are respectively the relativistic and the Newtonian prediction for the evolution of the angular velocity of the pulsar, seen from Earth. 

%Since, as we said in section \ref{approccioperturbativosullag}, the typical lags are characterised by roughly 
%\begin{equation}
%\Omega_{vp} = (1-\epsilon_n)\Omega_{np} \approx const,
%\end{equation}
%we obtain that $\Omega_{np}$ does not depend on $\theta$ (we recall that $\epsilon_n = \epsilon_n(r)$ in slow rotation approximation). Thus the angles $\theta$ at which we find the rings which transfer most of the angular momentum are around $\theta=\pi/2$ (equatorial plane), because it is the angle which maximizes the inertia moment of the ring. Finally, employing the fact that the metric functions vary slowly in the crust and can, therefore, be evaluated simply in $r=R_D$, and using the RP approximated formula \eqref{mistermuscoloidraulicogel} for $\mathfrak{N}_Z$ on the equatorial plane we get
%\begin{equation}\label{FfFormula}
%\dfrac{t_{R,\textrm{GR}}}{t_{R,\textrm{Newt}}} = \dfrac{e^{-\Phi_D}}{\lambda_D e^{-\Lambda_D}},
%\end{equation}
%where $t_{R,\textrm{GR}}$ is the relativistic prediction for the rise-time, while $t_{R,\textrm{Newt}}$ is the Newtonian one. 

We can now discuss the physical interpretation of equation \eqref{dododavecchio} by taking into account all the expected relativistic effects:
%\begin{equation}
%\dfrac{t_{R,\textrm{GR}}}{t_{R,\textrm{Newt}}} = e^{-\Phi_D} \cdot e^{\Lambda_D} \cdot \dfrac{1}{e^{-\Phi_D}} \cdot \dfrac{1}{1-\tilde{\omega}_D}.
%\end{equation}  
%Each of the four factors we isolated in the foregoing equation has its own interpretation and could be guessed a priori using simple arguments:
\begin{itemize}
	
\item \textit{Gravitational time dilation}: from Earth we observe a slow-motion picture of the internal dynamics of the star, thus the rise-time should be increased by a factor $e^{-\Phi_D}$. On the other hand, the time dilation increases the amount of vortices because an observer sitting inside the star sees faster motions. As the number of vortices increases, the mutual friction becomes stronger, so $t_{R,\textrm{GR}}$ is reduced by the factor $e^{\Phi_D}$. Thus the effects of time dilation cancel out.

\item \textit{Curvature of space}: in subsection \ref{approssimazionesulladensit} we have shown that the effect of gravity on the $t=const$ leaves is to enlarge the lengths in the radial direction. This reduces the density of vortices, increasing the rise-time by a factor $e^{\Lambda_D}$.

\item \textit{Frame dragging}: the frame-dragging constitutes a constant which needs to be subtracted to the angular velocity to connect what we observe from the Earth to what we would see if we were inside the star. Thus, it is not directly involved in the dynamical process. However it reduces the number of vortices, giving a factor $1-\tilde{\omega}_D$ in the denominator. 
\end{itemize}

We can rewrite \eqref{dododavecchio} in a more convenient way, considering that we can approximate the evaluation of the metric functions at the drip point with an evaluation at the surface of the star. In particular, given that
\begin{equation}
e^{\Lambda_D} \approx \bigg( 1-\dfrac{2M}{R} \bigg)^{-1/2} \spc  \tilde{\omega}_D \approx \dfrac{2I}{R^3},
\end{equation}
where $M$ is the Komar mass and $I$ is the Hartle moment of inertia of the whole star, we arrive at
\begin{equation}\label{IlGranFinale}
\dfrac{t_{R,\textrm{GR}}}{t_{R,\textrm{Newt}}} = \bigg( 1-\dfrac{2M}{R} \bigg)^{-1/2} \bigg( 1 - \dfrac{2I}{R^3} \bigg)^{-1}.
\end{equation}
This formula can be written only in terms of the relativistic compactness $\mathfrak{C}=M/R$ by using the universal relation 
\begin{equation}\label{Universal}
\dfrac{I}{R^3} = \bar{a}_1 \mathfrak{C}^2 + \bar{a}_2 \mathfrak{C} + \bar{a}_3 +\bar{a}_4 \mathfrak{C}^{-1} \, ,
\end{equation}
where  $\bar{a}_1 = 8.134 \times 10^{-1}$, $\bar{a}_2 = 2.101 \times 10^{-1}$, $\bar{a}_3 = 3.175 \times 10^{-3}$ and $\bar{a}_4 =-2.717 \times 10^{-4}$ are coefficients which do not depend of the EOS  \citep{Rezzolla_Universal_2016}.
%\begin{center}
%\begin{tabular}{|c|c|c|c|} \hline {$10\,\bar{a}_1$} & {$10\,\bar{a}_2$} & {$\bar{a}_3$} & {$\bar{a}_4$}\\ \hline
%$8.134 $ & $2.101  $ & $3.175 \times 10^{-3}$ & $-2.717 \times 10^{-4}$\\ \hline
%\end{tabular}
%\end{center}
%The final formula is, therefore,
%\begin{equation}
%\dfrac{t_{R,\textrm{GR}}}{t_{R,\textrm{Newt}}} =-\dfrac{1}{2} \bigg( 1-2\mathfrak{C} \bigg)^{-1/2} \bigg(   \bar{a}_1 \mathfrak{C}^2 + \bar{a}_2 \mathfrak{C}+\bar{a}_3 + \bar{a}_4 \mathfrak{C}^{-1}  -\dfrac{1}{2}\bigg)^{-1}.
%\end{equation}
Since all the formulas are given in geometric units, $\mathfrak{C}$, $I/R^3$ and the $\bar{a}_i$ are dimensionless.

\begin{figure}
\begin{center}
	\includegraphics[width=0.5\textwidth]{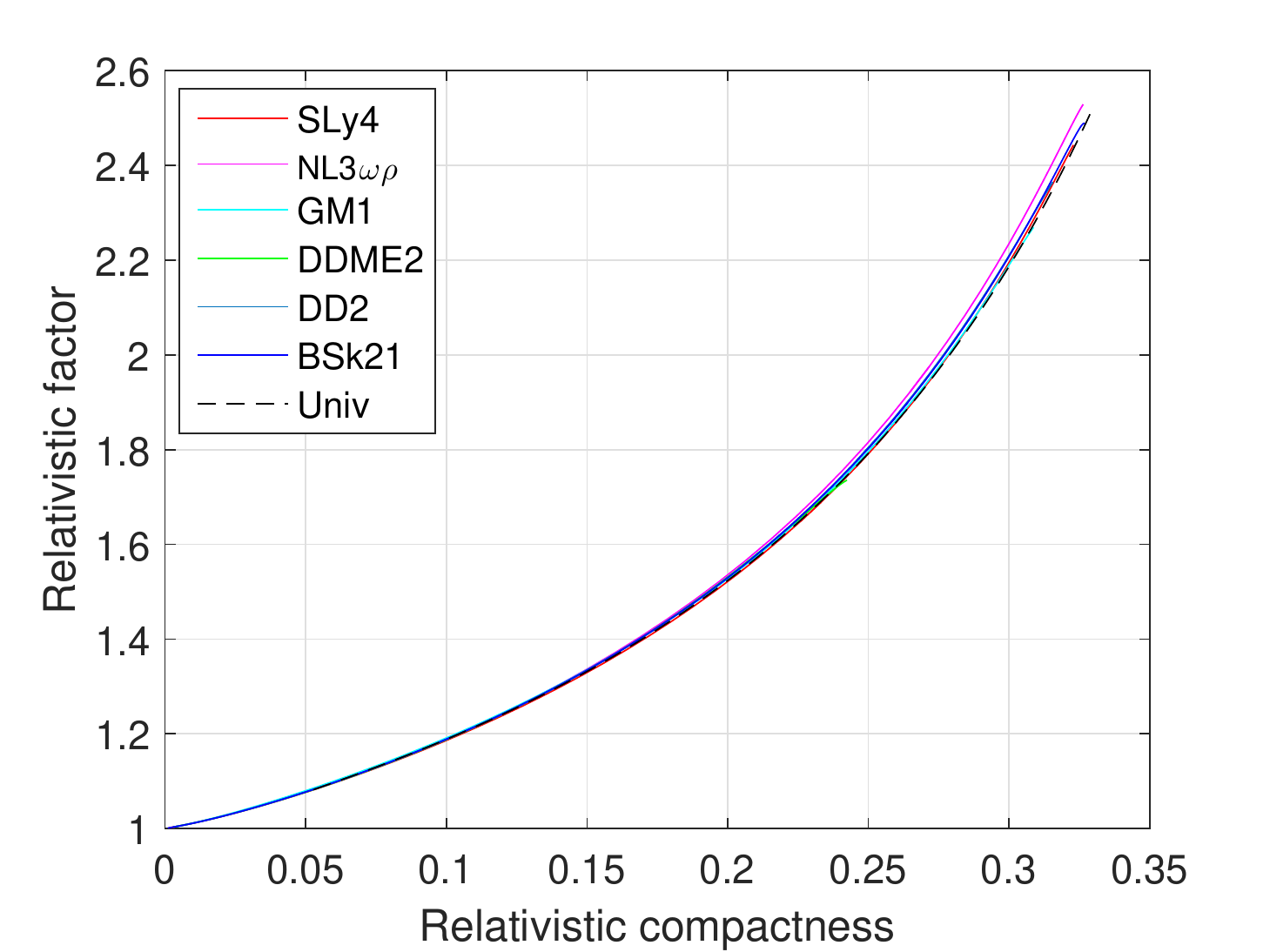}
	\caption{
		The relativistic factor $t_{R,\textrm{GR}}/t_{R,\textrm{Newt}}$ given in equation \eqref{IlGranFinale} as a function of the compactness $\mathfrak{C}$. 
		For comparison purposes, we employed six different equations of state \citep{Fortin2016,Fortin2017} to compute the moment of inertia $I=I(\mathfrak{C})$: all the curves appear to be nearly superimposed to the universal relation \eqref{Universal}. }
	\label{fig:risetimeccorrection}
	\end{center}
\end{figure}

In figure \ref{fig:risetimeccorrection} we show the relativistic correction as a function of the compactness: the curve in the plot is practically independent from the equation of state used. 
We observe that the relativistic correction is always larger than 1 and for a typical star of 1.4 $M_\odot$ ($\mathfrak{C} \approx 0.18$) it is around $1.4$. This means that the coupling time-scale are always longer.

%This is the general formula for the relativistic correction to the glitch rise-time, which you can see plotted in figure \ref{fig:risetimeccorrection} as a function of the mass for two different equations of state. As you can see it is always larger than one, meaning that the effect of GR is a slowing down of the process, and increases as relativity becomes more important. Notice also that the two plots diverge from each other as the mass increases.

\section{The role of the thin-shell assumption}\label{CompSour}

In this final section we discuss why the thin-shell assumption is fundamental to find a universal formula for the relativistic correction to the mutual friction coupling time-scale.

An alternative approach to obtain a global correction is the one adopted by \cite{sourie_glitch2017}, where both the species are assumed to move rigidly. 
%and work with an averaged version of equation \eqref{parmuUU}. In this framework it is impossible to appreciate the dependence on the latitude of the coupling time-scale showed in equation \eqref{dipendeontheta}. 
Differently from our approach (based on the presence of a thin shell that contains the superfluid neutrons), \cite{sourie_glitch2017} focus on the relaxation process in the case in which the free superfluid component is extended over the whole core of the star. 
This introduces two complications. First, the quantities of interest are obtained as integrals over the star, loosing the universal character of equation \eqref{IlGranFinale} (which has been found by evaluating all the metric functions at the surface). 
Secondly, the frame dragging $\omega$ is a function of both $\Omega_p$ and $\Omega_n$, correcting the coefficient $1-\epsilon_n$ in equation \eqref{hhhhhhhh} with a Lense-Thirring contribution (see the discussion in section \ref{the end}). This is another effect which compromises the universality of the result, as it strongly depends on the details of internal stratification.

In this section we start from a rigid model and we impose the requirement that the density of the neutron superfluid, $n_n$, (which is formally extended over the whole core) is zero outside a thin shell located near the surface. We prove that under this assumption the formula for the relativistic correction to the coupling time-scale given by \cite{sourie_glitch2017}  reduces to \eqref{dododavecchio}, while in general it may differ from it.

%In this final section we compare our results with those of \cite{sourie_glitch2017}. In the first four subsections we show analytically, in four steps, how their formula for the correction to the glitch rise-time reduces to \eqref{dododavecchio}. Subsection \ref{QA} is, then, dedicated to a qualitative comparison between figure 8 of \cite{sourie_glitch2017} and our quantitative predictions.

\subsection{Moments of inertia}

Following \cite{sourie_glitch2017} we have to introduce the partial moments of inertia of each species, because they appear directly in their formula for the correction to the coupling time-scale. As we said in subsection \ref{SMLapprox}, in the thin shell limit the metric is essentially unaffected by the presence of the lag, implying that
\begin{equation}\label{approxxxXXxx}
\omega = \Omega_p \tilde{\omega}.
\end{equation}
Using the slow rotation approximation, one can verify that the moment of inertia of the species $X$ defined in \cite{sourie_glitch2017} is
\begin{equation}
\hat{I}_X = \int_\Sigma m_n n_X (1-\tilde{\omega})e^{\Lambda + \Phi_D - 2\Phi} r^2 \sin^2 \theta \, d_3 x, 
\end{equation}
where $\Sigma$ is a $t=const$ hypersurface, $d_3x =r^2 \sin \theta \, dr \, d\theta \, d \varphi$, is the Newtonian volume element, and we have employed equation \eqref{jJjJjJj} to replace $\mu_X^{(C)}$. 
In the case $X=n$ the integrand is different from zero only in the thin shell, so the metric functions are approximately constant in the integral and can be replaced by their value on the drip point, thus we find
\begin{equation}
\hat{I}_n =(1-\tilde{\omega}_D)e^{\Lambda_D-\Phi_D}\int_\Sigma m_n n_n  r^2 \sin^2 \theta \, d_3 x. 
\end{equation}
On the other hand, for $X=p$, the integral is extended over the whole star, so $\hat{I}_p \gg \hat{I}_n$. Defining $\hat{I}= \hat{I}_p+\hat{I}_n$, we have that
\begin{equation}\label{zucche}
\hat{I}_p/\hat{I} \approx 1.
\end{equation}

\subsection{The gravitational space dilation factor}

A second quantity which appears in the formula for the relativistic correction of \cite{sourie_glitch2017} is the factor
\begin{equation}
\zeta := \dfrac{1}{2\hat{I}_n\Omega_n} \int_\Sigma \Gamma_{nZ} \, n_n k \mathfrak{N}_Z \, {\perp}(\partial_\varphi, \partial_\varphi) e^\Lambda d_3 x.
\end{equation} 
Using the slow rotation approximation, we have that
\begin{equation}
\Gamma_{nZ} \approx 1  \spc {\perp}(\partial_\varphi, \partial_\varphi) \approx r^2 \sin^2 \theta.
\end{equation}
Furthermore, since in the integral the angular part is weighed with a factor $\sin^3 \theta$, we can replace $k\mathfrak{N}_Z$ with its value in $\theta = \pi/2$, which, using \eqref{mistermuscoloidraulicogel}, reduces to
\begin{equation}
k\mathfrak{N}_Z = 2 m_n \Omega_n \lambda e^{-\Lambda}.
\end{equation}
The thin shell approximation allows to replace all the metric functions in the integral with their value in $R_D$; recalling equation \eqref{lambdinopiccolo}, we obtain
\begin{equation}\label{prigles}
\zeta \approx e^{-\Lambda_D}.
\end{equation}
Therefore $\zeta$ contains the correction to the density of vortices given by the gravitational dilation of space in the radial direction.

\subsection{The generalised entrainment coefficients}

In \cite{sourie_glitch2017}, the role of the entrainment and of the frame dragging on the coupling time-scale is encoded in two \textit{generalised entrainment coefficients} $\hat{\epsilon}_p$ and $\hat{\epsilon}_n$. In this subsection we recap how they are defined and how they simplify under the thin-shell assumption.
 
Given and arbitrary function $f$, we introduce the two averaging procedures ($X=n,p$)
\begin{equation}\label{Avvocato}
\braket{f}_X := \dfrac{\int_\Sigma f \, m_n n_X e^{\Lambda +\Phi_D -2\Phi } r^2 \sin^2 \theta \, d_3 x}{\int_\Sigma m_n n_X e^{\Lambda +\Phi_D -2\Phi } r^2 \sin^2 \theta \, d_3 x}.
\end{equation}
In particular,  \cite{sourie_glitch2017}  define
\begin{equation}
\tilde{\epsilon}_X = \braket{\epsilon_X}_X .
\end{equation}
It is clear that $\epsilon_p = 0$ outside the superfluid domain. Since the denominator (an integral over the whole star) is much larger than the numerator (restricted over the thin shell), we have that
\begin{equation}\label{BbuSs}
\tilde{\epsilon}_p \approx 0 \, .
\end{equation}
The authors also average the frame dragging and split the contributions as 
\begin{equation}
\braket{\omega}_X = \epsilon^{LT}_{XX} \Omega_X + \epsilon^{LT}_{YX} \Omega_Y.
\end{equation}
Using the approximation \eqref{approxxxXXxx}, it is clear that
\begin{equation}\label{TtoNinNo!!}
\epsilon^{LT}_{nn} \approx \epsilon^{LT}_{np} \approx 0  \spc  \epsilon^{LT}_{pn} \approx \tilde{\omega}_D.
\end{equation}
Finally they define the coefficients
\begin{equation}
\hat{\epsilon}_X = \dfrac{\tilde{\epsilon}_X - \epsilon^{LT}_{YX}}{1-\epsilon^{LT}_{XX}-\epsilon^{LT}_{YX}},
\end{equation}
which, considering \eqref{BbuSs} and \eqref{TtoNinNo!!}, become
\begin{equation}\label{cestinare}
\hat{\epsilon}_p \approx 0   \spc  \hat{\epsilon}_n \approx \dfrac{\tilde{\epsilon}_n -\tilde{\omega}_D}{1-\tilde{\omega}_D}.
\end{equation}

\subsection{Relativistic correction to the rise time}

Now we have all the ingredients we need to study the relativistic correction to the rise time of \cite{sourie_glitch2017}, 
\begin{equation}\label{cometichiami}
\dfrac{t_{R,\textrm{GR}}}{t_{R,\textrm{Newt}}} = \dfrac{\hat{I}_p^{\textrm{GR}}/\hat{I}^{\textrm{GR}}}{\hat{I}_p^{\textrm{Newt}}/\hat{I}^{\textrm{Newt}}} \cdot \dfrac{1-\hat{\epsilon}_p^{\textrm{GR}}-\hat{\epsilon}_n^{\textrm{GR}}}{1-\tilde{\epsilon}_p^{\textrm{Newt}}-\tilde{\epsilon}_n^{\textrm{Newt}}} \cdot \zeta^{-1} \, ,
\end{equation}
in the thin-shell approximation. 

Noting that equations \eqref{zucche} and \eqref{BbuSs} are true both in Newtonian theory and in General Relativity, we can use \eqref{prigles} and \eqref{cestinare} to obtain
\begin{equation}
\dfrac{t_{R,\textrm{GR}}}{t_{R,\textrm{Newt}}} =  \dfrac{1-\tilde{\epsilon}_n^{\textrm{GR}}}{1-\tilde{\epsilon}_n^{\textrm{Newt}}} \cdot \dfrac{e^{\Lambda_D}}{1-\tilde{\omega}_D} \, .
\end{equation}
However,  $\tilde{\epsilon}_n^{GR} \approx \tilde{\epsilon}_n^{\textrm{Newt}}$, because the relativistic corrections (given by the metric functions) can be brought out of the average over the thin shell, canceling out. 
%Clearly, this would not be possible if the integral was not extended over the whole star.
Thus, we finally arrive at
\begin{equation}\label{xsxsx}
\dfrac{t_{R,\textrm{GR}}}{t_{R,\textrm{Newt}}} =   \dfrac{e^{\Lambda_D}}{1-\tilde{\omega}_D} \, ,
\end{equation}
which is what we wanted to prove.

If the thin shell approximation is not valid, however, \eqref{zucche}, \eqref{BbuSs} and \eqref{xsxsx} do not hold and the relativistic correction will depend on the EOS as is seen in figure 8 of \cite{sourie_glitch2017}.

To date, there is no general consensus on the real extension of the superfluid region involved in the rise of a glitch. The standard scenario of a pinned superfluid confined only in the inner crust has been challenged in \cite{andersson+2012}, \cite{chamel2013} and \cite{pizzochero2019core}. An extension of the region including only an external fraction of the core for young neutron stars has been proposed by \cite{Ho_2015}. This model remains in the limit of the thin shell assumption. On the other hand, \cite{Alpar2014} have proposed that the whole core may contribute to the glitch, which would imply the need of employing the scheme adopted by \cite{sourie_glitch2017}.

%
%\subsection{Quantitative analysis}\label{QA}
%
%We have shown that at a mathematical level our formula is consistent with \cite{sourie_glitch2017}. Thus, the quantitative difference between our results must be a product of the fact that in their model the reservoir of angular momentum is assumed to be in the whole core, while in our case it is restricted to a thin superficial shell. In particular, the discrepancy we find is the product of two contributions: the ratio of the moments of inertia and the average of the entrainment coefficients. 
%
%The first is the result of the fact the moments of inertia are modified by relativity, so the ratio $\hat{I}_p/\hat{I}$ changes considering the relativistic corrections. However in our model this contribution is irrelevant because this ratio is essentially 1. 
%
%Secondly, one has to consider that the averaging procedure \eqref{Avvocato} contains some metric functions which in Newtonian must be removed. Thus different parts of the star have a different weight in the average going from the Newtonian theory to the Einsteinian one. Of course this does not happen in our case because for $X=n$ the metric functions can be brought out and for $X=p$ the results can be approximated to zero. Referring to figure 8 of \cite{sourie_glitch2017} the conclusion is that in our case the yellow line is always identically 0 and the green line coincides with the red one.
%

\section{Conclusions}\label{sec:conclusions}

We have computed the relativistic correction to the coupling time-scale between the superfluid and the normal component in a neutron star.

In doing this, we analysed all the quantities involved in the mutual friction equation. First, we studied the vortex profile in general relativity, verifying that the formula presented in \cite{Rothen1981} can be easily justified directly from Carter's two-fluid formalism in the slow rotation approximation. 
Secondly, we found that the validity of the approximations presented in \cite{Ravenhall_Pethick1994} implies that the vortex lines are expected to be almost straight in the quasi-Schwarzschild coordinates. 
Furthermore, we derived a formula for the vortex density, which, making use of the approximations of \cite{Ravenhall_Pethick1994}, becomes proportional to a geometric factor encoding the gravitational dilation of space in the radial direction.

Inserting all the results in the prescription for the vortex-mediated mutual friction presented in \cite{langlois98} we have shown that the relativistic corrections to the coupling time-scale are given, in the crust and in the outermost part of the outer core, by a universal factor which is a function of the compactness of the star only. 
This universal correction incorporates the effects of space curvature, which dilates the vortex spacing, and of the frame dragging, which reduces the amount of vortices. 
Both these effect reduce the mutual friction between the two species, slowing down the coupling process. For a typical star of 1.4 $M_\odot$ the glitch rise-time is enhanced of the $40\%$ with respect to Newtonian predictions. The correction grows as the star becomes more compact (and thus relativistic). 

Currently, Newtonian models are mostly employed to fit glitch rise-times and extract constraints on the mutual friction coefficients \citep{Haskell2018Crab, ashton2019rotational, pizzochero2019core}, with the notable exception of \citep{sourie_glitch2017}, who however consider the superfluid reservoir to be in the core of the star. 
In the case in which the reservoir is assumed to be in the crust, the mutual friction coefficients obtained by means of Newtonian models should just be rescaled with the coefficient given in equation \eqref{IlGranFinale} to encode the corrections of General Relativity. 
This result is particularly useful because the factor is a universal function of the compactness and, therefore, does not depend on the equation of state. We remark that \cite{Ho_2015} and \cite{NewtonBergerHaskell2015} propose techniques to distinguish equations of state based on differences that have the same order of magnitude as the relativistic correction that we obtain. Thus in these studies the effect of General Relativity cannot be neglected and our prescription for the correction needs to be adopted.

The general picture emerging from the present paper is that, despite the intrinsic difficulties of a fully relativistic approach, in the quasi-Schwarzschild coordinates all the relativistic effects assume a simple and intuitive form. 
The factors $e^\Phi$ (encoding gravitational time dilation), $e^\Lambda$ (encoding the curvature of space) and $\tilde{\omega}$ (encoding the Lense-Thirring effect) always appear in positions which are coherent with their intuitive meaning and their presence could also be deduced by means of simple arguments (see subsections \ref{Kubo1}, \ref{approssimazionesulladensit} and \ref{Kubo2}). 

Finally, the present work (together with similar ones, see e.g. \cite{langlois98}, \cite{andersson_comer2000}, \cite{sourie_glitch2017} and \cite{antonelli+2018}) can be used as a theoretical basis for the development of refined relativistic dynamical models for pulsar glitches and neutron star oscillations in the framework of the slow rotation approximation.

\section*{Acknowledgements}

The authors thank the PHAROS COST Action (CA16214) for partial financial support.
The authors acknowledge support from the Polish National
Science Centre grant SONATA BIS 2015/18/E/ST9/00577 and OPUS 2019/33/B/ST9/00942.
We thank A. Montoli and M. Fortin for valuable help.

\appendix

\section{Trough the eyes of an ideal observer}\label{misurabile}

Let us consider an ideal observer $\mathcal{O}$ with a four-velocity $u_{\mathcal{O}}$. 
We define the pseudovorticity associated to $\mathcal{O}$ as
\begin{equation}\label{gigi}
\varpi_{\mathcal{O}}^\nu := - {\star}\varpi^{\nu \rho} \, u_{\mathcal{O} \rho} = \varpi \, \mathcal{S}^{\nu \rho} u_{\mathcal{O} \rho}.
\end{equation} 
Considering the antisymmetry of ${\star}\varpi^{\nu \rho}$, it is evident that
\begin{equation}\label{gimli}
\varpi_{\mathcal{O}}^\nu u\indices{_{\mathcal{O}\nu}} =0,
\end{equation}
so that in the local reference frame of the observer it is a spatial vector. 
Equation \eqref{gandalf} immediately gives
\begin{equation}
{\perp}\indices{^\nu _\rho} \varpi_{\mathcal{O}}^\rho= 0, 
\end{equation}
which tells us that the pseudovorticity is tangent to the wordlsheet. 
Hence, $\varpi_{\mathcal{O}}$ points along the intersection between the vortex worldsheet and the local set of simultaneous events of the observer (this intersection is just the profile of the vortex locally seen by $\mathcal{O}$). 
Furthermore, we define
\begin{equation}\label{gollum}
v_{\mathcal{O}} := \dfrac{\varpi_{\mathcal{O}} }{\sqrt{g(\varpi_{\mathcal{O}},\varpi_{\mathcal{O}})}}, 
\end{equation}
which is the unit vector which points along the vorticity lines seen by the observer. Using \eqref{frodo}, it is easy to show that the denominator in the definition \eqref{gollum} can be rewritten as
\begin{equation}\label{aragorn}
\sqrt{g(\varpi_{\mathcal{O}},\varpi_{\mathcal{O}})} = \varpi \sqrt{-g(\paral u_{\mathcal{O}},\paral u_{\mathcal{O}})}.
\end{equation}
Another useful vector is
\begin{equation}\label{pralna}
u_{{V\mathcal{O}}} := \dfrac{\paral u_{\mathcal{O}}}{\sqrt{-g(\paral u_{\mathcal{O}},\paral u_{\mathcal{O}})}},
\end{equation}
which we refer to as \emph{vortex four-velocity} (with respect to the observer $\mathcal{O}$). 
Equations \eqref{frodo} and \eqref{aragorn} can be used to verify that 
\begin{equation}\label{tosto}
u_{V{\mathcal{O}}}^\nu = \mathcal{S}^{\nu \rho} v_{{\mathcal{O}}\rho},
\end{equation}
implying
\begin{equation}\label{legolas}
u_{V{\mathcal{O}}}^\nu v\indices{_{\mathcal{O}} _\nu}=0.
\end{equation}
To understand the meaning of $u_{V{\mathcal{O}}}$, imagine to mark a point in the core of a vortex. The worldline of this point is contained into the wordsheet of the corresponding vortex, implying that its four-velocity will be tangent to it. Clearly, there are infinitely many ways this particle can slide along the vortex line and this is reflected in the fact that the kernel of $\varpi_{\nu \rho}$ is a two-dimensional plane containing several normalised future-oriented timelike four-vectors $u_{C}$. In general, these $u_{C}$ are solutions of \eqref{saruman} and can be written as
\begin{equation}
u_{C}(\zeta) := u_{V{\mathcal{O}}} \cosh(\zeta) + v_{\mathcal{O}}\sinh(\zeta).
\end{equation}
Let us impose that the marked point is moving with four-velocity $u_{V{\mathcal{O}}}$, which is the case $\zeta =0$. Then,
\begin{equation}\label{brillantino}
\Gamma_{V{\mathcal{O}}} := -g(u_{\mathcal{O}},u_{V{\mathcal{O}}}) = \sqrt{-g(\paral u_{\mathcal{O}},\paral u_{\mathcal{O}})} 
\end{equation} 
is the Lorentz factor associated with its relative speed with respect to $O$, while the vector
\begin{equation}\label{3velocity}
w_{V{\mathcal{O}}} := \dfrac{u_{V{\mathcal{O}}}}{\Gamma_{V{\mathcal{O}}}} - u_{\mathcal{O}},
\end{equation}
represents the three-velocity of the particle seen by the observer. Moreover  $w_{V{\mathcal{O}}}$ is orthogonal (making use of \eqref{gimli} and \eqref{legolas}) to $v_{\mathcal{O}}$. 
This means that $u_{V{\mathcal{O}}}$ represents the four-velocity that the matter contained in the core of the vortices would have if we suppose that in the frame of $\mathcal{O}$ its motion is perpendicular to the shape of the vortices. 
For this reason, it can be considered to be the four-velocity of the vortices with respect to $\mathcal{O}$, as it describes how their profile moves in their local frame. 
In particular, it is easy to show that
\begin{equation}
\Gamma_{V{\mathcal{O}}} = \dfrac{1}{\sqrt{1-\Delta_{V{\mathcal{O}}}^2}} \spc  \Delta_{V{\mathcal{O}}}^2 = g(w_{V{\mathcal{O}}},w_{V{\mathcal{O}}}).
\end{equation}
%In appendix \eqref{sec:drift} we show that, in an electromagnetic analogue, the three-velocity $w_{V\mathcal{O}}$ is nothing but the $E \times B$ drift velocity, see \cite{bellan_2006}.   
%

We now derive the local density of vortices measured by the observer. 
To do so, it is necessary to use the Feynman-Onsager quantization condition which emerges from \eqref{gringo}, see also \cite{antonelli+2018}. 
Consider a spacelike two-dimensional surface $\Sigma$ in the superfluid domain, enclosed by a circuit $\partial \Sigma$. It is immediate to see that
\begin{equation}\label{motta}
\int_{\partial \Sigma} \mu = k\mathcal{N} ,
\end{equation}
where $\mathcal{N}$ is the net number of vortices enclosed by $\partial \Sigma$ \citep{antonelli+2018}. 
Employing Stokes' theorem and the fact that the exterior derivative commutes with the pull-back,
\begin{equation}
\mathcal{N} = \dfrac{1}{k} \int_{\Sigma} \varpi.
\end{equation}
Let us define at each point of $\Sigma$ an orthonormal basis $e_a$, such that $e_0$ (time-like) and $e_1$ are normal to it, while $e_2$ and $e_3$ are tangent. In the text we employ the convention that, in the integral, the pull-back of $\varpi$ is identified with $\varpi$ itself. Since, however, the pull-back of $\varpi$ is a 2-form defined in a two-dimensional space, then 
\begin{equation}
\int_{\Sigma} \varpi= \int_{\Sigma} \varpi(e_2,e_3)\sqrt{G} \, d_2 x,
\end{equation}
where $\sqrt{G}$ is the square root of the determinant of the metric induced on $\Sigma$. 
Considering the fact that in a Lorentzian manifold the Hodge dual has the property
\begin{equation}
{\star}{\star} = -1
\end{equation}
on two-forms, we can write
\begin{equation}
\varpi (e_2,e_3) = \varpi_{23} = -\varepsilon_{0123} {\star}\varpi^{01} = {\star}\varpi(e_0,e_1).
\end{equation}
Defining the unit bivector normal to the surface
\begin{equation}
\mathcal{S}^{\nu \rho}_{\Sigma} := e_0^\nu e_1^\rho - e_0^\rho e_1^\nu,
\end{equation}
it is evident that
\begin{equation}
 {\star}\varpi(e_0,e_1) = \dfrac{1}{2} {\star}\varpi_{\nu \rho} \mathcal{S}^{\nu \rho}_{\Sigma} \, .
\end{equation}
Plugging our result into the integral we finally obtain
\begin{equation}
\mathcal{N} =\dfrac{1}{2k} \int_{\Sigma} {\star}\varpi_{\nu \rho} d\Sigma^{\nu \rho},
\end{equation}
where we used the compact notation
\begin{equation}
d\Sigma^{\nu \rho} := \mathcal{S}^{\nu \rho}_{\Sigma} d\Sigma   \spc \text{with} \spc  d\Sigma:=\sqrt{G} \, d_2 x.
\end{equation}
Therefore, the integral of the two-form has been transformed into the flux of its Hodge dual. Defined the scalar
\begin{equation}\label{giochiamocela}
\mathfrak{N}:= \dfrac{\varpi}{k},
\end{equation}
we are able to write
\begin{equation}
\mathcal{N} = -\dfrac{1}{2} \int_{\Sigma} \mathfrak{N} \mathcal{S}_{\nu \rho}  d\Sigma^{\nu \rho}.
\end{equation}
The density of vortices is a number per unit area and must be computed by considering surfaces which are orthogonal to the profile of the vortices. Thus, to obtain the local density of vortices $\mathfrak{N}_{\mathcal{O}}$ measured by $\mathcal{O}$, we need to consider an infinitesimal area $\delta S$ that is orthogonal to both $u_{\mathcal{O}}$ and $v_{\mathcal{O}}$.
This gives
\begin{equation}
\mathcal{S}^{\nu \rho}_{\Sigma} = u_{\mathcal{O}}^\nu v_{\mathcal{O}}^\rho - u_{\mathcal{O}}^\rho v_{\mathcal{O}}^\nu 
\end{equation}
and using \eqref{pralna}, \eqref{tosto} and \eqref{brillantino}, we arrive at
\begin{equation}\label{zutturale}
\mathfrak{N}_{\mathcal{O}} = \dfrac{\mathcal{N}}{\delta \Sigma} = - \mathfrak{N} \mathcal{S}^{\nu \rho} u_{{\mathcal{O}}\nu} v_{{\mathcal{O}}\rho} = \mathfrak{N} \Gamma_{V{\mathcal{O}}}.
\end{equation}
Looking again at equation \eqref{gollum} and \eqref{aragorn}, we notice that the norm of $\varpi_{\mathcal{O}}$ is the density measured by $O$ (apart from the factor $k$). Hence, the pseudovorticity can be written as 
\begin{equation}
\varpi_{\mathcal{O}} = k \mathfrak{N}_{\mathcal{O}} v_{\mathcal{O}}.
\end{equation}
Notice that if the four-velocity is tangent to the wordsheet (meaning that for the relative observer the vortices are locally at rest), the density of vortices $\mathfrak{N}_{\mathcal{O}}$ reduces to $\mathfrak{N}$. This tells us that $\mathfrak{N}$ can be considered to be the \emph{rest-frame vortex density}. Now the formula \eqref{bilbo} has a clear interpretation: the factor $\Gamma_{V{\mathcal{O}}}$ encodes the contraction of lengths, which modifies the densities only if the velocity of the observer has a component which is orthogonal to the vortex profile.

\section{Relativistic mutual friction}\label{sec:Reduction of the mutual friction}

Let us start by rearranging the terms of \eqref{mutualfriction2} as
\begin{equation}\label{orasiamobrutti}
( g_{\nu \rho} - \mathcal{R}^{-1}\Gamma_{vp}^{-1}\varepsilon_{\nu \mu \rho \sigma} u_n^\mu  v_p^\sigma)u_v^\rho = \Gamma_{vp}^{-1} u_{p\nu}. 
\end{equation}
This relation defines $u_v$ in terms of $u_p$, $v_p$ and $u_n$: we can solve it for $u_v$, considering now $\Gamma_{vp}$ as a parameter and ignoring for the moment the fact that it is a function of $u_v$ itself. 
To simplify the calculations we choose a convenient orthonormal basis $e_a = e_a^\mu \partial_\mu$ such that $e_0 = u_n$ and $v_p \in span \{ e_0 , e_1 \}$, with $v_p^1>0$. 
In this tetrad equation \eqref{orasiamobrutti} reads
\begin{equation}\label{orasiambelli}
\mathcal{M}_{ab} u_{v}^b = \Gamma_{vp}^{-1} u_{pa},
\end{equation}
where
\begin{equation}
\mathcal{M}_{ab}= \eta_{ab} + \mathcal{Z} \, \varepsilon_{01ab} \spc  \mathcal{Z}= \mathcal{R}^{-1} \Gamma_{vp}^{-1} v_p^1,
\end{equation}
whose matrix representation on this basis is
\begin{equation}\label{Mmmm}
\mathcal{M}_{ab} =
\begin{bmatrix}
  -1  & 0 &  0 &  0  \\
   0 & 1 & 0 & 0  \\
   0 & 0 & 1 & \mathcal{Z}  \\
   0 & 0 & -\mathcal{Z} & 1
\end{bmatrix}.
\end{equation} 
Now,  if we are able to find the tensor $\mathcal{M}^{ca}_{(inv)}$ such that
\begin{equation}
\mathcal{M}^{ca}_{(inv)} \mathcal{M}_{ab} = \delta\indices{^c_b},
\end{equation}
then contracting \eqref{orasiambelli} with $\mathcal{M}^{ca}_{(inv)}$ gives
\begin{equation}\label{cisiarrivaeasy}
u_v^c = \mathcal{M}_{(inv)}^{ca} \Gamma_{vp}^{-1}u_{pa},
\end{equation}
which is the expression we are looking for. 
Since the matrix form of $\mathcal{M}_{(inv)}$ is the inverse of $\mathcal{M}$, it is now clear why the choice of this basis is so convenient: in fact, it is easy to invert \eqref{Mmmm}, namely 
\begin{equation}
\mathcal{M}^{ca}_{(inv)} =
\begin{bmatrix}
  -1  & 0 &  0 &  0  \\
   0 & 1 & 0 & 0  \\
   0 & 0 & \dfrac{1}{1+\mathcal{Z}^2} & -\dfrac{\mathcal{Z}}{1+\mathcal{Z}^2}  \\
   0 & 0 & \dfrac{\mathcal{Z}}{1+\mathcal{Z}^2} & \dfrac{1}{1+\mathcal{Z}^2}
\end{bmatrix},
\end{equation}
which in a tensorial notation can be written as
\begin{equation}\label{chepotenzalaGR}
\mathcal{M}_{(inv)}^{ca} = \eta^{ca} - \dfrac{\mathcal{Z}}{1+ \mathcal{Z}^2} \varepsilon\indices{_0_1^c^a} - \dfrac{\mathcal{Z}^2}{1+\mathcal{Z}^2} \hat{\perp}^{ca},
\end{equation}
where $\hat{\perp}\indices{^\nu _\rho}$ is the projector onto the plane generated by $e_2$ and $e_3$.\footnote{Not to be confused with $\perp\indices{^\nu _\rho}$ of equation \eqref{frodo}}

We use \eqref{chepotenzalaGR} into \eqref{cisiarrivaeasy}, lower the index $c$ and rewrite everything in a generic coordinate basis, obtaining
\begin{equation}\label{velocitamedullaesatta}
u_{v\nu} = \dfrac{u_{p\nu}}{\Gamma_{vp}} - \dfrac{\mathcal{R}^{-1}}{1+\mathcal{Z}^2} u_n^\lambda v_p^\sigma \varepsilon_{\lambda \sigma \nu \rho} \dfrac{u_p^\rho}{\Gamma_{vp}^2} - \dfrac{\mathcal{Z}^2}{1+\mathcal{Z}^2} \hat{\perp}_{\nu \rho} \dfrac{u_p^\rho}{\Gamma_{vp}}.
\end{equation}
This equation is an exact reformulation of \eqref{mutualfriction}. In our model, however, we will work in the limit in which the relative speeds between all the species are non relativistic, so we can neglect the Lorentz factors and consider the limit
\begin{equation}
\mathcal{Z} \longrightarrow \dfrac{1}{\mathcal{R}}.
\end{equation}
It is, also, useful to introduce the relative three-velocities, defined in general by the condition
\begin{equation}
u_x =: \Gamma_{xy}(u_y + w_{xy}),
\end{equation}
which in this limit become
\begin{equation}
w_{xy}=u_x - u_y.
\end{equation}
Equation \eqref{velocitamedullaesatta}, then, reduces to
\begin{equation}\label{velocitvortici}
w_{vp\nu} = \dfrac{\mathcal{R}}{1+\mathcal{R}^2} u_n^\lambda v_p^\sigma \varepsilon_{\lambda \sigma \nu \rho} w_{np}^\rho + \dfrac{\hat{\perp}_{\nu \rho}w_{np}^\rho}{1+\mathcal{R}^2} .
\end{equation}
Consider again equation \eqref{mutualfriction}; in the limit of small relative speeds
\begin{equation}
{\perp} u_p \approx -w_{vp} 
\end{equation}
and we can rewrite the right-hand side employing \eqref{velocitvortici}, arriving at
\begin{equation}
-u_n^\rho \varpi^n_{\rho \nu} = \dfrac{\mathcal{R}k\mathfrak{N}}{1+\mathcal{R}^2} \bigg[  \mathcal{R} u_n^\lambda v_p^\sigma \varepsilon_{\lambda \sigma \nu \rho} w_{np}^\rho + \hat{\perp}_{\nu \rho}w_{np}^\rho  \bigg],
\end{equation}
which is the expression of mutual friction we were looking for (see \citealt{Andersson_Mutual_Friction_2016} for an alternative derivation).

\section{Closed degenerate two-forms in General Relativity}\label{CDTFIGRSPTM}

In this appendix we briefly review the geometrical properties of two-forms in General Relativity, expanding some concepts that have been touched in section \ref{superfluids}. In particular, we show that the degeneracy condition \eqref{buzz} is sufficient to guarantee the existence of a two-dimensional foliation whose leaves are tangent to the kernel of the two-form. 
In the following, the names of the tensors which we introduce are those of the fields of electrodynamics in General Relativity. This because the formalism we are going to build is employed exactly as presented here also in GRMHD \citep{special_in_gen,Gralla_jacobson2014}.

Consider an arbitrary two-form $F$. Given the four-velocity $u_{\mathcal{O}}$ of an observer, we can build a right-handed orthonormal basis $e_a=e_a^\mu \partial_\mu$ such that $e_0 =u_{\mathcal{O}}$. The form $F$ can be expanded in this basis as
\begin{equation}
F = \dfrac{1}{2} F_{ab} \, e^a \wedge e^b 
\end{equation}
and the components $F_{ab}$ represent what is measured by the observer moving with $u_{\mathcal{O}}$. In this basis we can rename the components of  $F_{ab}$ according to the Faraday decomposition
\begin{equation}
F_{ab} = 
\begin{bmatrix}
  0  & -E_1 &  -E_2 &  -E_3  \\
   E_1 & 0 & B_3 & -B_2  \\
   E_2 & -B_3 & 0 & B_1  \\
   E_3 & B_2 & -B_1 & 0
\end{bmatrix}.
\end{equation}
We can introduce the two covectors
\begin{equation}
E_{\mathcal{O}} := E_j e^j  \spc  B_{\mathcal{O}} := B_j e^j,
\end{equation}
where $j$ runs from $1$ to $3$ and we put the subscript ${\mathcal{O}}$ to keep track of the fact that everything depends on the choice of the observer. Using the musical duality notation
\begin{equation}
u_{\mathcal{O}}^\flat := u_{{\mathcal{O}}\nu} dx^\nu ,
\end{equation}
$F$ can be rewritten as
\begin{equation}\label{impressioanante_1}
F = u_{\mathcal{O}}^\flat \wedge E_{\mathcal{O}} + \star(u_{\mathcal{O}}^\flat \wedge B_{\mathcal{O}}).
\end{equation}
Moreover, 
\begin{equation}
(\star F)_{ab} = 
\begin{bmatrix}
  0  & B_1 &  B_2 &  B_3  \\
   -B_1 & 0 & E_3 & -E_2  \\
   -B_2 & -E_3  & 0 & E_1  \\
   -B_3 & E_2 & -E_1 & 0
\end{bmatrix},
\end{equation}
which can be seen as the result of the transformation 
\begin{equation}
E_{\mathcal{O}} \rightarrow -B_{\mathcal{O}}   \spc  B_{\mathcal{O}} \rightarrow E_{\mathcal{O}} 
\end{equation}
and implying that
\begin{equation}\label{impressioanante_2}
\star F = - u_{\mathcal{O}}^\flat \wedge B_{\mathcal{O}} + \star(u_{\mathcal{O}}^\flat \wedge E_{\mathcal{O}}).
\end{equation}
The properties of the wedge product and of the Hodge duality, together with  equations \eqref{impressioanante_1} and \eqref{impressioanante_2}, can be used to prove that
\begin{equation}\label{impressioanante_3}
E_{\mathcal{O}} = - \iota_{u_{\mathcal{O}}} F  \spc  \spc  B_{\mathcal{O}} =  \iota_{u_{\mathcal{O}}} {\star}F.
\end{equation}
Thus we have a simple way to extract, for a given $F$ and $u_{\mathcal{O}}$, the two covectors $E_{\mathcal{O}}$ and $B_{\mathcal{O}}$. Notice also that $E_{\mathcal{O}}$ and $B_{\mathcal{O}}$ can be combined to give two scalars which do not depend on the choice of $u_{\mathcal{O}}$. In fact, it is immediate to verify that
\begin{equation}\label{equestapasserallastoria}
\begin{split}
& \braket{F,F} = \braket{B_{\mathcal{O}},B_{\mathcal{O}}}-\braket{E_{\mathcal{O}},E_{\mathcal{O}}}  \\  
&\braket{F,{\star}F} = 2\braket{E_{\mathcal{O}},B_{\mathcal{O}}},
\end{split}
\end{equation}
where $\braket{,}$ is the inner product of forms. 
Now, let us suppose there is a four-velocity field $u_C$ such that
\begin{equation}\label{mnbvcx}
\iota_{u_C} F =0
\end{equation}
everywhere. According to \eqref{impressioanante_3} this is equivalent to requiring that in every point of the spacetime the quantity $E$ associated to $u_C$ is equal to zero. We call (employing again the musical duality formalism)
\begin{equation}
\mathbb{B} := (\iota_{u_C} {\star}F)^\#,
\end{equation} 
which is the $B$ associated to $u_C$ with raised indices. Hence, from \eqref{impressioanante_1}, we immediately have that 
\begin{equation}\label{topologicamenteimmenso}
F = \star(u_C^\flat \wedge \mathbb{B}^\flat).
\end{equation}
Now in every point of the spacetime we can consider the plane $\mathcal{K}:= span \{u_C,\mathbb{B} \}$. This two-dimensional plane coincides with the kernel of $F$, i.e. it is the set of all the vectors $v$ such that $\iota_v F=0$. 
According to Frobenius' theorem, the condition for the existence of a regular foliation of the spacetime in two-dimensional worldsheets which are in every point tangent to the corresponding $\mathcal{K}$ is that for any couple of vector fields $X$ and $Y$ which take values in $\mathcal{K}$ everywhere, their commutator takes values in $\mathcal{K}$ itself. However, from
\begin{equation}
\iota_{[X,Y]} F = [\mathcal{L}_X , \iota_Y] F,
\end{equation}
and using Cartan's magic formula and the conditions $\iota_X F = \iota_Y F=0$, one finds that the condition for the hypotheses of the Frobenius theorem to hold is that 
\begin{equation}
\iota_X \iota_Y dF=0,
\end{equation}
for all $X$, $Y$ in $\mathcal{K}$. In particular, if $dF=0$, the existence of the foliation is guaranteed. 

In the electromagnetic setting $F=dA$, so $dF=d^2A=0$ and the degeneracy condition is verified in the case of force-free GRMHD \citep{Gralla_jacobson2014}. The worldsheet foliation, then, describes evolution of the magnetic-field lines. 

As a last remark, note that equation \eqref{mnbvcx} automatically implies, according to \eqref{equestapasserallastoria}, that
\begin{equation}
\braket{F,{\star}F}=0,
\end{equation}
which is equivalent to
\begin{equation}
F \wedge F =0.
\end{equation}

\subsection{$E \times B$ drift velocity}\label{sec:drift}

Consider a closed and degenerate two-form $F$ which can be written in the form presented in equation \eqref{topologicamenteimmenso}. Then, for a given observer with four-velocity $u_\mathcal{O}$ it is possible to define the quantity
\begin{equation}\label{sonoalcamk}
w_{E\times B}^{\mathcal{O}} :=  \dfrac{\star (u_\mathcal{O} \wedge E_\mathcal{O} \wedge B_\mathcal{O})}{\braket{B_\mathcal{O},B_\mathcal{O}}}.
\end{equation}
In the frame of the observer this four-vector takes the form
\begin{equation}
{w_{E\times B}^{\mathcal{O}}}^j = \dfrac{(E_\mathcal{O} \times B_\mathcal{O})^j}{|B_\mathcal{O}|^2}.
\end{equation} 
If $F$ is the Faraday tensor, so we can interpret $E_\mathcal{O}$ and $B_\mathcal{O}$ as respectively the electric and magnetic fields in the frame of $\mathcal{O}$, then $w_{E\times B}^{\mathcal{O}}$ is the so called $E \times B$ drift velocity \citep{bellan_2006}. 

Let us write $w_{E\times B}^{\mathcal{O}}$ in terms of $u_C$, $\mathbb{B}$ and $u_\mathcal{O}$. 
It is convenient to work in components; combining \eqref{impressioanante_3} and \eqref{topologicamenteimmenso}, we find that
\begin{equation}\label{orsi}
\begin{split}
& E_{\mathcal{O}\mu} = \varepsilon_{\mu \nu \rho \sigma} u_{\mathcal{O}}^\nu u_C^\rho \mathbb{B}^\sigma \\
& B_{\mathcal{O}\mu} = \Gamma_{\mathcal{O}C} \mathbb{B}_\mu + g(u_\mathcal{O},\mathbb{B}) u_{C\mu},
\end{split}
\end{equation}
which immediately implies
\begin{equation}
\braket{B_\mathcal{O},B_\mathcal{O}} =  g(\mathbb{B},\mathbb{B})\Gamma_{\mathcal{O}C}^2 - g(u_\mathcal{O},\mathbb{B})^2.
\end{equation}
By definition, the numerator of the right-hand side of equation \eqref{sonoalcamk} is, in components,
\begin{equation}
\star (u_\mathcal{O} \wedge E_\mathcal{O} \wedge B_\mathcal{O})^\mu = \varepsilon^{\nu \rho \sigma \mu} u_{\mathcal{O}\nu} E_{\mathcal{O}\rho} B_{\mathcal{O}\sigma}.
\end{equation}
Plugging equations \eqref{orsi} inside this identity, and using the condition
\begin{equation}
\varepsilon^{\lambda \nu_1 \nu_2 \nu_3} \varepsilon_{\lambda \rho_1 \rho_2 \rho_3} = -3! \delta\indices{^{[\nu_1} _{\rho_1}} \delta\indices{^{\nu_2} _{\rho_2}} \delta\indices{^{\nu_3]} _{\rho_3}},
\end{equation}
it is possible to verify that
\begin{equation}
\begin{split}
 \star (u_\mathcal{O} \wedge E_\mathcal{O} \wedge B_\mathcal{O})^\mu = & 
 g(\mathbb{B},\mathbb{B}) \Gamma_{\mathcal{O}C} u_C^\mu + g(u_\mathcal{O},\mathbb{B}) \mathbb{B}^\mu \\ 
 & - [\Gamma_{\mathcal{OC}}^2 g(\mathbb{B},\mathbb{B})-g(u_\mathcal{O}, \mathbb{B})^2 ] u_{\mathcal{O}}^\mu.
\end{split}
\end{equation}
Finally, considering that 
\begin{equation}
\paral \indices{^\nu _\rho} = - u_C^\nu u_{C\rho} + \dfrac{\mathbb{B}^\nu \mathbb{B}_\rho}{g(\mathbb{B},\mathbb{B})},
\end{equation}
so that
\begin{equation}
(\paral u_\mathcal{O})^\mu = \Gamma_{\mathcal{O}C} u_C^\mu + \dfrac{g(u_\mathcal{O},\mathbb{B})}{g(\mathbb{B} , \mathbb{B})} \mathbb{B}^\mu
\end{equation}
and
\begin{equation}
- g(\paral u_\mathcal{O},\paral u_\mathcal{O}) =  \Gamma_{\mathcal{O}C}^2 - \dfrac{g(u_\mathcal{O},\mathbb{B})^2}{g(\mathbb{B},\mathbb{B})},
\end{equation}
we arrive at the formula
\begin{equation}
w_{E\times B}^{\mathcal{O}} = \dfrac{\paral u_\mathcal{O}}{- g(\paral u_\mathcal{O},\paral u_\mathcal{O})} - u_\mathcal{O},
\end{equation}
which is the analogue of the three-velocity given in equation \eqref{3velocity}.

\section{Derivation using the language of forms}\label{deriv with forms}

In this appendix we provide a  formal derivation of the results in section \ref{saacc} by using the language of the exterior calculus.

Our starting point is equation  \eqref{ilmomentone}. The exterior derivative of this formula gives the four-vorticity two-form which, using the properties of the exterior derivative, can be written as
\begin{equation}\label{leomardo}
\varpi = d\mu_t \wedge dt + d\mu_\varphi \wedge d\varphi \, .
\end{equation}
It is easy to check that this is equivalent to \eqref{immaturo}. We have shown in appendix \ref{CDTFIGRSPTM} that a two-form  must satisfy the condition
\begin{equation}
\varpi \wedge \varpi =0
\end{equation} 
to be degenerate (i.e. to have a non-trivial kernel).
More explicitly, this means that
\begin{equation}
d\mu_t \wedge d\mu_\varphi \wedge dt \wedge d\varphi =0.
\end{equation}
This is a requirement of linear dependence which, considering that $d\mu_t$ and $d\mu_\varphi$ are linear combinations only of $dx$ and $dy$, is equivalent to
\begin{equation}
d\mu_t \wedge d\mu_\varphi =0.
\end{equation}
If we assume that $d\mu_\varphi$ is nowhere zero, we can recast the above constrain as the requirement that there exists a function $\Omega_C$ such that
\begin{equation}\label{michelangeloeraffaello}
d\mu_t = -\Omega_C d\mu_\varphi .
\end{equation}
Let us imagine to build a coordinate system such that the function $\mathcal{N}$ is one of the coordinates. Then, the above equation, which in an arbitrary chart reads
\begin{equation}
\partial_\nu \mu_t = -\dfrac{k\Omega_C}{2\pi} \partial_\nu \mathcal{N},
\end{equation}
in this system of coordinates becomes
\begin{equation}
\partial_\nu \mu_t =  -\dfrac{k\Omega_C}{2\pi} \delta\indices{^{\mathcal{N}} _\nu}.
\end{equation}
This implies
\begin{equation}
\mu_t = \mu_t (\mathcal{N})
\end{equation}
and 
\begin{equation}
\dfrac{d\mu_t}{d\mathcal{N}} = -\dfrac{k\Omega_C}{2\pi} ,
\end{equation}
namely
\begin{equation}
\Omega_C = \Omega_C (\mathcal{N}).
\end{equation}
This completes the formal proof for the validity of equation \eqref{chepwer!}. Let us put equation \eqref{michelangeloeraffaello} into \eqref{leomardo}, as well as the definition of $\mathcal{N}$:  we arrive at 
\begin{equation}
\varpi = \dfrac{k}{2\pi} d\mathcal{N} \wedge (d\varphi - \Omega_C dt), 
\end{equation}
that provides  a simpler and more compact way of writing \eqref{maturo}. Furthermore, it is  transparent that the vector $u_C$ defined in \eqref{pulzella} belongs to the kernel of the four-vorticity.

We can take advantage of the exterior calculus to find also the vortex rest-frame density. 
In fact, we know that 
\begin{equation}
\braket{\varpi, \varpi } = k^2 \mathfrak{N}^2 \, .
\end{equation}
On the other hand,  if $\alpha^{(1)}, ... ,\alpha^{(p)}, \beta^{(1)},...,\beta^{(p)}$ are one-forms, then
\begin{equation}
\braket{\alpha^{(1)} \wedge ... \wedge \alpha^{(p)}, \beta^{(1)}\wedge .... \wedge \beta^{(p)}} = \det [\braket{\alpha^{(j)},\beta^{(K)}}] \, .
\end{equation}
Therefore, considering that $d\mathcal{N}$ and $d\varpi - \Omega_C dt$ are manifestly orthogonal, we find that
\begin{equation}
\mathfrak{N}^2 = \dfrac{1}{4 \pi^2}\braket{d\mathcal{N}, d \mathcal{N}} \braket{d\varphi - \Omega_C dt,d\varphi - \Omega_C dt}.
\end{equation}
The first scalar product is  
\begin{equation}
\braket{d\mathcal{N}, d \mathcal{N}} = (g_x^{-1}\partial_x \mathcal{N})^2 + (g_y^{-1}\partial_y \mathcal{N})^2,
\end{equation}
while  it is easy to verify that
\begin{equation}
\braket{d\varphi - \Omega_C dt,d\varphi - \Omega_C dt} = \dfrac{1}{\rho^2 \Gamma_{CZ}^2},
\end{equation}
which leads us to the final result
\begin{equation}
\mathfrak{N} =\dfrac{\sqrt{ (g_x^{-1}\partial_x \mathcal{N})^2 + (g_y^{-1}\partial_y \mathcal{N})^2  }}{2\pi \rho \Gamma_{CZ}} \, ,
\end{equation}
in agreement with what we found in equations \eqref{zzzzz} and \eqref{zzzzzzzzzzzzzzzzzzzzzzz}.

\bibliographystyle{mnras}
\bibliography{Biblio}

\bsp
\label{lastpage}
\end{document}